\documentclass[12pt,a4paper]{article}
\pdfoutput=1
\usepackage{jheppub}
\usepackage[T1]{fontenc} 
\usepackage{epsfig} 
\usepackage{amscd}
\usepackage{verbatim}
\usepackage{latexsym}
\usepackage{amsfonts,amsthm,amsmath,mathtools}
\usepackage{amssymb,stmaryrd,textcomp,enumerate,bbm}
\usepackage{color}
\usepackage{xcolor}
\usepackage{booktabs}
\usepackage{float}
\usepackage{graphicx}
\usepackage{lscape}
\usepackage{slashed}

\notoc 

\title{\centering{Surveying 4d SCFTs\\ twisted on Riemann surfaces }}
\author[a]{Antonio Amariti}
\author[b]{\!\!,~Luca Cassia}
\author[b]{\!\!,~Silvia Penati}
\affiliation[a]{Albert Einstein Center for Fundamental Physics, Institute for Theoretical Physics, University of Bern,Sidlerstrasse 5, Bern, ch-3012, Switzerland}
\affiliation[b]{Universit\`a degli studi di Milano Bicocca and INFN, Sezione di Milano--Bicocca, Piazza della Scienza 3, 20161, Milano, Italy}
\emailAdd{amariti@itp.unibe.ch,luca.cassia@mib.infn.it,silvia.penati@mib.infn.it}

\newcommand{\g}{{\gamma}}

\newcommand{\mathe}{\mathrm{e}}
\newcommand{\mathi}{\mathrm{i}}

\newcommand{\nocomma}{}
\newcommand{\tmmathbf}[1]{\ensuremath{\boldsymbol{#1}}}
\newcommand{\tmop}[1]{\ensuremath{\operatorname{#1}}}

\newcommand{\nocontentsline}[3]{}
\newcommand{\tocless}[2]{\bgroup\let\addcontentsline=\nocontentsline#1{#2}\egroup}

\abstract{
Within the framework of four dimensional conformal supergravity we consider $\mathcal{N}=1,2,3,4$ supersymmetric theories generally twisted along the abelian subgroups of the R--symmetry and possibly other global symmetry groups. Upon compactification on constant curvature Riemann surfaces with arbitrary genus we provide an extensive classification of the resulting two dimensional theories according to the amount of supersymmetry that is preserved. 
Exploiting the c--extremization prescription introduced in arXiv:1211.4030 we develop a general procedure to obtain the central charge for 2d $\mathcal{N}=(0,2)$ theories and the expression of the corresponding R--current in terms of the original 4d one and its mixing with the other abelian global currents.
}

\begin{document}

\maketitle

\newpage
\tableofcontents

\section{Introduction}
Two dimensional (super) conformal field theories ((S)CFTs) play a central role in the
worldsheet description of string theory and in the formulation of the AdS$_3$/CFT$_2$ correspondence.
Moreover, being the conformal group infinite dimensional, many 
exact results can be extracted from their algebraic structure.
Classifying 2d CFTs is anyway a difficult task and finding new examples of conformal theories 
is not straightforward.

A powerful laboratory to build infinite families of 2d CFTs is supersymmetry.
SCFTs in 2d can be obtained
by compactifying 4d SCFTs on curved compact 2d manifolds. 
In this process some of the original supersymmetry charges survive whenever Killing spinor equations 
arising from requiring fermion variations to vanish, admit non--trivial 
solutions. In general, this does not happen since on curved manifolds there are no covariantly constant Killing spinors. However, as suggested in  \cite{Witten:1988xj} (see also \cite{Bershadsky:1995vm,Bershadsky:1995qy}), 
this problem can be circumvented by performing a (partial) topological twist, i.e. by turning on background gauge fields for (a subgroup of) the R--symmetry group
along the internal manifold in such a way that its contribution to the Killing spinor equations compensates the contribution from the spin connection.  
More generally, one can also turn on properly quantized background fluxes for other non--R flavor symmetries. In this case preserving supersymmetry also requires to set to zero the associated gaugino variations.
 
Although this procedure does not allow to extract the matter content of the 2d theory, useful information on its IR behavior is provided by 
the 2d global anomalies that can be obtained in terms of the 4d ones and of the background fluxes \cite{Benini:2012cz}.

Focusing on 2d theories with $\mathcal{N}=(0,2)$  (or equivalently, $\mathcal{N}=(2,0)$) supersymmetry, the corresponding central charge $c_{L}$ ($c_R$) is proportional to the anomaly of the abelian R--symmetry current inherited from the exact 4d R--current $J_R^\mu$, obtained by $a$-maximization \cite{Intriligator:2003jj}. However, under dimensional reduction 4d abelian global currents can mix with the exact 4d $J_R^\mu$, hence the exact 2d R--current has to be determined by extremizing the 2d central charge $c_{L}$ ($c_R$) as a function of such a mixing.   
The program of constructing $\mathcal{N}=(0,2)$  2d SCFTs from 4d became an intense field of research 
\cite{Benini:2013cda,Karndumri:2013iqa,Kutasov:2013ffl,Kutasov:2014hha,Bobev:2014jva,Bea:2015fja,Gadde:2015wta,Benini:2015bwz,Apruzzi:2016nfr,Amariti:2016mnz,Hosseini:2016cyf,Lawrie:2016axq}
after such c--extremization principle was derived in \cite{Benini:2012cz}.
 
An interesting phenomenon regarding the mixing of global currents with $J_R^\mu$  
has been observed in \cite{Benini:2015bwz} for the particular case of 4d $\mathcal{N}=1$ $Y^{pq}$ quiver theories compactified on Riemann surfaces. 
There, it was observed that even though there is a global (baryonic) symmetry, that does not mix with $J_R^\mu$ at the 4d fixed point \cite{Bertolini:2004xf,Butti:2005vn}, 
it has a non--trivial mixing with $J_R^\mu$ at the 2d fixed point. This phenomenon is generalizable to cases with a richer structure of baryonic symmetries.
\\

Motivated by the former discussion, in this paper we engineer the partial topological twist in the natural setup of conformal supergravity and study
systematically the twisted compactification on constant curvature Riemann surfaces of 4d SCFTs with different amount of supersymmetry.
In this unified framework we investigate the cases of $\mathcal{N}=1,2,3,4$ conformal supergravity corresponding to 4d geometries of the form $\mathbbm{R}^{1, 1} \times \Sigma$ where $\Sigma$  is a genus $g$ Riemann surface. We study the conditions
to preserve different amounts of supersymmetry in 2d by solving the Killing spinor equations arising from setting to zero the variations of the gravitino and of the auxiliary fermions in the Weyl multiplet (Sections \ref{sec:N=1}, \ref{sec:N=2}, \ref{sec:N=3} and  \ref{sec:N=4}).  
When possible (i.e. in cases with $\mathcal{N}=1,2$ supersymmetry) we also turn on vector multiplets associated to global non--R symmetries.
In this case an additional constraining equation for Killing spinors arises from setting to zero the variation of the corresponding gaugino. 

All possible cases are listed in Tables \ref{TableN1} ($\mathcal{N}=1$), \ref{TableN2} ($\mathcal{N}=2$), \ref{TableN3} ($\mathcal{N}=3$) and \ref{TableN4} ($\mathcal{N}=4$).
In $\mathcal{N} =1,2$ cases the presence of global gauged non--R
symmetries in general decreases (or does not increase) the number of supersymmetries, but never below $\mathcal{N} = (0,2)$ or $(2,0)$.
For $\mathcal{N}=3,4$ theories, where flavor symmetries are absent, we also discuss the possibility of twisting in two steps.
This consists in a first twist along an abelian subgroup of $SU(3)_R \times U(1)_R$ or $SU(4)_R$, reducing the R--symmetry and leaving some vector multiplets associated to  non--R global symmetries. 
A further twist along such symmetries corresponds to $\mathcal{N}=1$
or $\mathcal{N}=2$ gaugings and preserves half of the supercharges.

In section \ref{sec:central}, when the resulting 2d theories have  $\mathcal{N}=(0,2)$ supersymmetry, we provide 
the (formal) expression for the anomaly coefficients and the central charge  
in terms of the 4d anomalies and the fluxes, as obtained by c--extremization.  Concurrently, the explicit expression for the exact 2d R--current
is given as a linear combination of the 4d R--current and global non--R symmetries.
In section \ref{sec:conclusions} we conclude by commenting on some possible future lines of research.
In appendix \ref{app:A} few necessary details on the anomaly polynomial are collected.
In appendix \ref{app:B} we provide further details on the vanishing of the supersymmetry variation for the auxiliary fermions in the $\mathcal{N}=3,4$ cases. 

%
%
%
%
%
%
\section{Twisted reduction of $\mathcal{N}=1$ SCFTs}
\label{sec:N=1}
%
%
%
%
%
%
%
We begin by considering a $\mathcal{N}= 1$ superconformal theory on the four dimensional spin manifold $M = \mathbbm{R}^{1, 1} \times \Sigma$, 
where $\Sigma$ is a Riemann surface of genus $g$ and constant scalar curvature. Twisted compactification of this class of theories has been already discussed in \cite{Johansen:2003hw,Gadde:2015wta,Benini:2015bwz}. Here we review the procedure in a $\mathcal{N}=1$ superconformal gravity setup to fix the general scheme that we will use in the $\mathcal{N}$--extended cases. 

$\mathbbm{R}^{1,1}$ coordinates are labelled $(x^0,x^1)$, while the ones on $\Sigma$ are $(x^2,x^3)$. The spin connection $\omega_{\mu}$ on $\Sigma$ satisfies the relation
\begin{equation}
 \label{eq:GaussBonnet}
 \frac{1}{2 \pi} \int_\Sigma R(\omega) = 2 - 2g
\end{equation}
where $\frac{R(\omega)}{2\pi}$ is precisely a representative 2--form for the first Chern class of the tangent bundle of $\Sigma$ expressed in terms of the Riemannian curvature $R(\omega)=d\omega$. Such a characteristic class is usually denoted as $c_1(T\Sigma)\in H^2(\Sigma,\mathbbm{Z})$.
The curvature for a Riemann surface can be written in terms of the volume 
form $d\tmop{Vol}_\Sigma$ and the Gaussian curvature $K$ as
\begin{equation}
 \label{eq:scalarcurvature}
 R(\omega) = K\,d\tmop{Vol}_\Sigma
\end{equation}
For later convenience we define the normalized scalar curvature $\kappa\equiv sgn(K)$, the
normalized volume form $\Omega_{\mu\nu}$
\begin{equation}
 \label{eq:volume}
 \Omega\equiv \left\{
   \begin{array}{lcl}
    |K|d\tmop{Vol}_\Sigma & \tmop{for} & K \neq 0\\
    ~ & ~ & ~ \\
    2\pi\frac{d\tmop{Vol}_\Sigma}{\tmop{Vol}_\Sigma} & \tmop{for} & K = 0
   \end{array} \right.
\end{equation}
and the total normalized volume $\nu$
\begin{equation}
 \label{eq:nu}
 \nu\equiv\int_\Sigma\frac{\Omega}{2\pi}
\end{equation}
so that $R(\omega)_{\mu\nu}=\kappa\,\Omega_{\mu\nu}$ and $\kappa\nu=2-2g$.
\\

In general, compactification on $\Sigma$ breaks supersymmetry completely, since on arbitrarily curved manifolds there are no covariantly constant Killing spinors. Along the lines of \cite{Festuccia:2011ws}, in order to put a 4d theory on a curved manifold and preserve some supersymmetry we couple the theory to a conformal supergravity background that reproduces the desired spacetime geometry. The whole superconformal group is gauged and the corresponding gauge fields are organized into the Weyl multiplet as follows (we use notations and conventions of \cite{book:931093}) 
\begin{table}[H]
\centering
 \label{tab:N=1weyl}
 \begin{tabular}{|c|ccccc|cc|}   
 \hline
 generator & $P_a$ & $M_{ab}$ & $\Delta$ & $K_a$ & $T_R$ & $Q$ & $S$ \\
  \hline
  field     & $e^a_\mu$ & $\omega_\mu^{ab}$  & $b_\mu$ & $f_\mu^a$ & $A_\mu$  & $\psi_\mu$ & $\phi_\mu$ \\
  \hline
 \end{tabular}
\caption{Generators and gauge fields of $\mathcal{N}=1$ conformal supergravity.}
\end{table}
\noindent
Here $P_a$,$K_a$ are vector generators of translations and special conformal transformations, $M_{ab}$
and $\Delta$ are generators of Lorentz rotations and dilatations, $Q$ and $S$ are the spinorial supercharges.
The $U(1)_R$ R--symmetry generator $T_R$ assigns charge $-1$ to the positive chirality supercharges $Q_\alpha$
and ${S}_{\alpha}$ and charge $+1$ to their conjugates
$\bar{Q}_{\dot{\alpha}}=(Q_\alpha)^\dagger$ and $\bar{S}_{\dot{\alpha}}=(S_\alpha)^\dagger$. When the R--symmetry generator acts on
the supercharges we will often write $T_R = -\g_5$ with $\g_{5}=\mathi\g_{0123}$.

The supersymmetry transformation laws of the independent gauge fields read
\begin{eqnarray}
 \label{eq:variations}
 \delta e_\mu^a  & = & \frac{1}{2}\bar{\varepsilon}\g^a\psi_\mu \\
 \delta b_\mu    & = & \frac{1}{2}\bar{\varepsilon}\phi_\mu - \frac{1}{2}\bar{\eta}\psi_\mu \\
 \delta A_\mu    & = & \frac{1}{2}\mathi\bar{\varepsilon}\g_5\phi_\mu + \frac{1}{2}\mathi\bar{\eta}\g_5\psi_\mu \\
 \delta \psi_\mu & = & \mathcal{D}_\mu\varepsilon - e^a_\mu\g_a\eta 
\label{eq:variationpsi}
\end{eqnarray}
where $\varepsilon, \eta$ are the Majorana spinors associated to $Q$ and $S$ transformations, respectively. The covariant derivative is defined as $\mathcal{D}_\mu\varepsilon \equiv (\partial_\mu + \frac{1}{2}b_\mu + \frac{1}{4}\omega_\mu^{ab} \g_{ab} - \mathi A_\mu T_R) \varepsilon$.  

Since we are only interested in theories on curved manifolds with rigid supersymmetry, we fix the Weyl multiplet to be a collection of background fields describing the geometry of spacetime. In order to preserve Lorentz invariance on $\mathbbm{R}^{1,1}$ we set all the spinor fields to zero and assign possibly non--vanishing components to bosonic forms only in the $(x^2, x^3)$ directions.  As follows from (\ref{eq:variationpsi}), in general this choice breaks superconformal invariance. However, some $Q$--supersymmetry survives if the geometry admits non--trivial covariantly constant spinor fields, solutions of the equation $\mathcal{D}_\mu\varepsilon = 0$ (setting $\eta = 0$) \footnote{To begin with one could solve the equation $\delta \psi_\mu = 0$ for non--vanishing $\eta$, by setting $\eta = \tfrac14 \slash {\hspace{-0.25cm} D}  \varepsilon$ \cite{Pestun:2014mja}. The solution $\eta=0, D_\mu \varepsilon = 0$ is compatible with this condition.}.  
This equation may have non--trivial solutions if we turn on a non--zero background also for the R--symmetry gauge connection $A_\mu$ \cite{Witten:1988xj} such that the two contributions coming from $A_\mu$ and $\omega_\mu^{ab}$ in the covariant derivative cancel each other. 

More precisely, focusing on constant solutions, we first apply the exterior derivative to $\delta\psi_\mu$,
so that the Killing spinor equation $\mathcal{D}_\mu\varepsilon = 0$  is traded with
\begin{equation}
 \label{eq:twistedKillingSpinorEq2}
 2\partial_{[\mu}\delta\psi_{\nu]} = \left[ \frac{1}{2} R_{\mu\nu}(\omega^{23})\g_{23}- \mathi R_{\mu\nu}(A)\g_5 \right] \varepsilon = 0 
\end{equation}
where $R_{\mu\nu}(\omega^{23})$ and $R_{\mu\nu}(A)$ are the curvatures of the connections $\omega^{23}_\mu$ and $A_\mu$, respectively.
Given the particular form of the curvature $R_{\mu\nu}(\omega)=\kappa\,\Omega_{\mu\nu}$, we choose $A_\mu$ such that its curvature is also proportional to the normalized volume form $\Omega_{\mu\nu}$
 \begin{equation}
 \label{eq:curvatureN=1}
 R_{\mu\nu}(A)=-a\,\Omega_{\mu\nu} 
\end{equation}
where the parameter $a$ is constrained by the Dirac quantization condition
\begin{equation}
 \label{eq:Dirac2}
 \frac{1}{2\pi} \int_\Sigma R(A) = -a\int_\Sigma\frac{\Omega}{2\pi} = -a \nu\in\mathbbm{Z} 
\end{equation}
Substituting (\ref{eq:curvatureN=1}) in (\ref{eq:twistedKillingSpinorEq2}), we then obtain
\begin{equation}
 \label{eq:twistedKillingSpinorEq3}
\left[\frac{\kappa}{2}\mathi\g_{23}-a\g_5\right]\varepsilon=0 
\end{equation}
We postpone the search and classification of non--vanishing solutions to section \ref{N=1solutions}{\color{blue}.2}.

%
%
%
%
%
%
\tocless\subsection{Twisting with flavors}\label{N=1k=0}
%
%
%
%
%
We now consider the case in which the original 4d theory also admits a global abelian non--R symmetry that can be either flavor or baryonic symmetry. With an abuse of notation, we call it $U(1)_{flavor}$. 

This symmetry can be weakly gauged by turning on a background connection \footnote{Similar discussions appeared in \cite{Almuhairi:2011ws,Benini:2013cda,Kutasov:2013ffl}.}.  However, in order to preserve the original superconformal symmetry one has to turn on a whole abelian $\mathcal{N}=1$ superconformal gauge multiplet $(B_\mu,\lambda,Y)$ whose field content consists of the gauge vector potential $B_\mu$, the gaugino $\lambda$ and the auxiliary scalar $Y$, all in the adjoint representation of the flavor symmetry. The corresponding supersymmetry transformations are
\begin{eqnarray}
 \label{eq:variationsGauge}
 \delta B_\mu & = & -\frac{1}{2}\bar{\varepsilon}\g_\mu\lambda  \nonumber\\
 \delta \lambda         & = &  \left[\frac{1}{4} \g^{ab} R_{ab}(B) +\frac{1}{2} Y \mathi\g_5\right]\varepsilon  \\
 \delta Y               & = &  \frac{1}{2}\mathi\bar{\varepsilon}\g_5\g^\mu\mathcal{D}_\mu\lambda \nonumber 
\end{eqnarray}
where $R_{\mu\nu}(B)$ is the curvature 2--form of the gauge connection $B_\mu$ and the covariant derivative on spinors is defined as in eq. (\ref{eq:variationpsi}). 

Similarly to the case of the R--symmetry background in (\ref{eq:curvatureN=1}), we can choose a $U(1)_{flavor}$ connection with constant curvature
\begin{equation}
 \label{eq:flavorN=1gauging}
 R_{\mu\nu}(B)=b\,\Omega_{\mu\nu} \quad , \quad b\nu\in\mathbbm{Z}
\end{equation}
together with vanishing background gaugino. 
In order to preserve some supersymmetry we have to require
\begin{equation}
 \label{eq:gaugino-variation}
 \delta\lambda = \left[\frac{b}{2}|e|\Omega_{23}\g_{23} + \frac{1}{2} Y \mathi\g_5\right]\varepsilon = 0 
\end{equation}
where $|e|=e^{22}e^{33}-e^{23}e^{32}$ is the vielbein determinant on $\Sigma$.

Writing $\g_5 = \mathi \g_{23} \g_{01}$ in the previous equation allows to factor out a gamma matrix $\g_{23}$. Therefore, setting $Y=\pm b |e|\Omega_{23}$ we finally obtain the condition
\begin{equation}
 \label{eq:flavor-eq-N=1}
 \left(1\mp\g_{01}\right)\varepsilon=0 
\end{equation}
We then see that in principle, turning on a background for an abelian non--R global symmetry, introduces additional constraints on the supersymmetry generators. 

More generally, we can consider 4d theories with rank--$n$ flavor symmetry group, i.e. with $n$
generators $T_i$ in the Cartan subalgebra. In this case we can gauge one vector multiplet $(B^i_\mu,\lambda^i,Y^i)$
for each Cartan generator.  If the corresponding
auxiliary scalars are fixed by the same equation $Y_i=+b_i|e|\Omega_{23}$ (or $Y_i=-b_i|e|\Omega_{23}$) we are led to the same constraints (\ref{eq:flavor-eq-N=1}).
\\

%
%
%
%
%
%
\tocless\subsection{Classification of the solutions}\label{N=1solutions}
%
%
%
%
%
We are now ready to discuss the most general solutions of the two supersymmetry preserving conditions
\begin{equation}
\left[\frac{\kappa}{2}\mathi\g_{23}-a\g_5\right]\varepsilon=0 \qquad , \qquad b_i  \left(1\mp\g_{01}\right)\varepsilon=0
\label{N=1constraints}
\end{equation}
where the constant $a$ signals the presence of a non--trivial $U(1)_R$ background, eq. (\ref{eq:curvatureN=1}),  while $b_i$ are associated to $B^i_\mu$ connections for $U(1)_{flavor}$ symmetries, eq. (\ref{eq:flavorN=1gauging}).  We note that the second equation is nothing but a 2d (anti)chirality condition. 

In order to find solutions to these equations, we write the Majorana spinor $\varepsilon$ in terms of its Weyl components, $\varepsilon = ( \epsilon_{\alpha} \; \, \bar{\epsilon}^{\dot{\alpha}})$, and with no loss of generality we restrict the discussion to the positive chiral spinor $\epsilon_\alpha$ transforming in the $\tmmathbf{2}$ of $SL(2, \mathbbm{C})$. 

On the product manifold $\mathbbm{R}^{1,1}\times\Sigma$ the original Lorentz group of 4d Minkowski is reduced as $Spin(3,1)\rightarrow Spin(1,1)\times Spin(2)_\Sigma$, and consequently the spinorial representation of $\epsilon_\alpha$ also splits as
\begin{equation}
 \label{eq:splittingN=1}
 \tmmathbf{2} \rightarrow \left[\tmmathbf{1}_{1,1}\oplus\tmmathbf{1}_{-1,-1}\right] 
\end{equation}
Here the representations on the right hand side are labelled by the eigenvalues of the hermitian generators $\g_{01}$ and $\mathi\g_{23}$
of $Spin(1,1)$ and $Spin(2)_\Sigma$, respectively.
The generator $\g_{01}$ corresponds also to the chirality operator on $\mathbbm{R}^{1,1}$, hence we refer to $\tmmathbf{1}_{1,1}$ and $\tmmathbf{1}_{-1,-1}$ as the 2d positive (left)  and negative (right) chirality representations respectively,  and denote the corresponding spinors as $\epsilon_+$  and $\epsilon_-$.

\begin{table}[h]
 \centering
  \begin{tabular}{|c|c|c|c|c|c|c|}
    \hline
    supersymmetry & chirality & representation & $\g_{01}$ & $\mathi \g_{23}$ & $\g_5$ & $\delta\psi_\mu=0$\\
    \hline
    $\epsilon_+$ & L & $\tmmathbf{1}_{+ 1, + 1}$ & $+ 1$ & $+ 1$ & $+1$ & $a - \kappa/2 = 0$\\
    $\epsilon_-$ & R & $\tmmathbf{1}_{- 1, - 1}$ & $- 1$ & $- 1$ & $+1$ & $a + \kappa/2 = 0$\\
    \hline
  \end{tabular}
  \caption{Supersymmetry generators and their charges under $\tmop{Spin} (1,
  1)$, $\tmop{Spin} (2)_{\Sigma}$ and R--symmetry. Since the $U(1)_R$ generator can be written as $\g_5=(\g_{01})(\mathi\g_{23})$, it follows that $\epsilon_\pm$ are automatically 
irreducible representations of the R--symmetry group corresponding to charge 1. 
}
 \label{tab:N=1}
\end{table}

As summarized in Table \ref{tab:N=1}, for $\kappa \neq 0$ solutions to the first eq. in (\ref{N=1constraints}) correspond to $\epsilon_+$ for $a=\frac{\kappa}{2}$ and $\epsilon_-$ for $a=-\frac{\kappa}{2}$. The second equation in (\ref{N=1constraints}) does not restrict the Killing spinors any further, since we can always choose $b_i$ such that (\ref{eq:flavor-eq-N=1}) projects on the same chirality as that of the Killing spinor.
Therefore, independently of the presence of gauged flavor symmetries, the resulting 2d theory is $\mathcal{N}=(2,0)$ for $a=\frac{\kappa}{2}$ and 
$\mathcal{N}=(0,2)$ for $a=-\tfrac{\kappa}{2}$. These solutions are compatible with the quantization condition $a \nu \in {\mathbbm{Z}}$, being $\kappa \nu$ an even number.

In the special case of compactification on a torus, $\kappa=0$, when no flavor symmetry is gauged ($b_i=0$) there is no need for twisting. In fact, setting $A_\mu$ to zero, the Killing spinor equation reduces to $ \partial_\mu \epsilon = 0$ and is automatically satisfied for every constant section $\epsilon$. Therefore, supersymmetry is not broken and the resulting 2d theory is $\mathcal{N}=(2,2)$ with R--symmetry $U(1)_{left}\times U(1)_{right}$ generated by the two combinations $T_{\pm} = \frac{1}{2}T_R \pm M_{23}$, where $M_{23}$ is the Lorentz generator on $\Sigma$. Supersymmetry can be reduced by gauging some flavor symmetry. In this case, in fact, the second equation in (\ref{N=1constraints}) constrains the supercharges to be of definite chirality and reduces supersymmetry to $\mathcal{N}=(2,0)$ for $Y_i=+ b_i |e| \Omega_{23}$ or $\mathcal{N}=(0,2)$ for $Y_i=-b_i |e| \Omega_{23}$.
 
The complete picture of topological twisted reduction of ${\mathcal N}=1$ SCFTs is summarized in Table
\ref{TableN1}, where the resulting 2d theories are classified in terms of the surviving amount of supersymmetry.
 \begin{table}[H]
\begin{center}
  \begin{tabular}{|c|cc|}   
  \hline
  $\kappa \neq 0$ & $a=\frac{\kappa}{2}$ & $a=-\frac{\kappa}{2}$ \\
  \hline
  $b = 0$ & $ \mathcal{N}=(2,0)$ & $\mathcal{N}=(0,2)$ \\
  $b \neq 0$ & $\mathcal{N}=(2,0)$ & $\mathcal{N}=(0,2)$ \\
  \hline
 \end{tabular}
~~
 \begin{tabular}{|c|c|}   
  \hline
  $\kappa=0$ & $a=0$ \\
  \hline
  $b = 0$ & $\mathcal{N}=(2,2)$ \\
  $b \neq 0$ & $\mathcal{N}=(2,0)$ or $(0,2)$ \\
  \hline
  \end{tabular}
\caption{Classification of topologically twisted 4d $\mathcal{N}=1$ SCFTs 
on Riemann surfaces of constant curvature $\kappa=\pm1,0$
in terms of the surviving amount of supersymmetry in 2d. We include 
the possibility of a twist along the flavor symmetries, with flux $b$. }
\label{TableN1}
\end{center}
\end{table}
\vskip 10pt
%
%
%
%
%
%
\section{Twisted reduction of $\mathcal{N}=2$ SCFTs}
\label{sec:N=2}
%
%
%
%
%
%
We now consider a $\mathcal{N}=2$ SCFT with R--symmetry group $SU(2)_R \times U(1)_R$.
The Lie algebra of $SU(2)_R$ is spanned by anti--hermitian matrices $\mathi\sigma_A$, where $\sigma_{A=1,2,3}$ are the three Pauli matrices.

The four--dimensional chiral supercharges $Q_{\alpha I}$ are in the $(\tmmathbf{2},\bar{\tmmathbf{2}})_{-1}$ representation
of the group $Spin(3,1) \times SU(2)_R \times U(1)_R$,
 while their complex conjugates $\bar{Q}_{\dot{\alpha}}^I=(Q_{\alpha I})^\dagger$ transform in the $({\bar{\tmmathbf{2}}},\tmmathbf{2})_{+1}$
 representation. In particular, the $U(1)_R$ generator $T_R$ acts on the supercharges as $-\g_5$.

The $\mathcal{N}=2$ superconformal algebra contains a $\mathcal{N}=1$ subalgebra with
R--symmetry group $U(1)_R^{\mathcal{N}=1}$ generated by the combination
\begin{equation}
 \label{eq:N=1subalgebra}
 T_R^{\mathcal{N}= 1} = \frac{2}{3} \sigma_{3} + \frac{1}{3} T_R 
\end{equation}

Twisted compactifications of $\mathcal{N}=2$ SCFTs have been already considered in \cite{Karlhede:1988ax,Kapustin:2006hi,Gadde:2015wta}. Here we give a systematic derivation within the superconformal gravity setup. 

Analogously to the $\mathcal{N}=1$ case, a $\mathcal{N}=2$ SCFT can be consistently defined on a curved manifold $M = \mathbbm{R}^{1, 1} \times \Sigma$, by first coupling 
it to the extended $\mathcal{N}=2$ superconformal gravity and then gauge fixing the background Weyl multiplet as to reproduce the desired geometry with possibly non--trivial fluxes turned on in order to preserve some supersymmetry. 

We recall that the $\mathcal{N}=2$ Weyl multiplet contains the gauge fields of the conformal group $e_\mu^a,f_\mu^a,b_\mu,\omega_\mu^{ab}$, the superconnections $\psi^I_\mu,\phi_{\mu I}$ associated to supersymmetries $Q_I$ and $S^I$, the connections $A_\mu$ and $V_{\mu}^A$ for the R--symmetry groups $U(1)_R$ and $SU(2)_R$ and the auxiliary fields $T_{ab}^-,D$ (bosonic) and $\chi^I$ (fermionic), needed to close the algebra off--shell.

Under supersymmetry transformations the fermionic fields of the gravity multiplet transform as
\begin{eqnarray}
 \label{eq:variationsN=2psi}
 \delta \psi_\mu^I & = & \left[\partial_\mu \!+\! \frac{1}{2}b_\mu \!+\! \frac{1}{4}\omega_\mu^{ab}\g_{ab}\!-\! A_\mu\mathi\g_5\right]\varepsilon^I \!-\!
                          V_{\mu}^A(\mathi\sigma_A)^I_J\varepsilon^J \!-\! \frac{1}{16}\g^{ab}T^-_{ab}\varepsilon^{IJ}\g_\mu\varepsilon_J \\
 \label{eq:variationsN=2chi}
 \delta \chi^I     & = & \frac{1}{2}D\varepsilon^I - \frac{1}{6}\g^{ab}\left[\frac{1}{4}\slashed{\mathcal{D}}T^-_{ab}\varepsilon^{IJ}\varepsilon_J
                           - R_{ab}(A)\mathi\g_5\varepsilon^I - R_{ab}(V^A)(\mathi\sigma_A)^I_J\varepsilon^J \right]
\end{eqnarray}

In order to preserve Lorentz invariance on $\mathbbm{R}^{1,1}$ the background fermions must be set to zero. This choice automatically sets to zero the $Q$--supersymmetry variation of all bosonic fields, which can then be chosen such that the $Q$--variation of the fermions vanish as well. 
 
From (\ref{eq:variationsN=2psi}) and (\ref{eq:variationsN=2chi}) we deduce that we can safely set the background fields $b_\mu$ and $T^-_{ab}$ to zero
and simplify these expressions to
\begin{eqnarray}
 \label{eq:variationsN=2psi2}
  \delta \psi_\mu^I & = & \left[\partial_\mu \!+\! \frac{1}{4}\omega_\mu^{ab}\g_{ab}\!-\! A_\mu\mathi\g_5\right]\varepsilon^I \!-\!
                          V_{\mu}^A(\mathi\sigma_A)^I_J\varepsilon^J  \equiv 0  \\
 \label{eq:variationsN=2chi2}
\delta \chi^I     & = & \frac{1}{2}D\varepsilon^I + \frac{1}{6}\g^{ab}\left[ R_{ab}(A)\mathi\g_5\varepsilon^I + R_{ab}(V^A)(\mathi\sigma_A)^I_J \varepsilon^J \right] \equiv 0
\end{eqnarray}
The remaining background connections $A_\mu$ and $V_\mu^A$  can then be used to perform partial topological twist as we now describe. 

Turning on a background flux for $V_\mu^A$ breaks explicitly the $SU(2)_R$ invariance of the theory down to a $U(1)$ subgroup of it. Without loss of generality we choose this subgroup to be the one generated by $\mathi\sigma_3$. Namely, we parametrize the R--symmetry gauging as follows
\begin{equation}
 \label{eq:N2Rsymmback}
 R_{\mu\nu}(A)= -a_1 \Omega_{\mu\nu}, \quad R_{\mu\nu}(V^{A=1,2})=0, \quad R_{\mu\nu}(V^{3})= -a_2 \Omega_{\mu\nu} 
\end{equation}
where the parameters $a_{i=1,2}$, are constrained by the quantization condition $a_i\nu\in\mathbbm{Z}$, and $\Omega_{\mu\nu}$ is the normalized volume form of $\Sigma$. This choice is actually equivalent to gauging the 1--parameter subgroup of $SU(2)_R\times U(1)_R$ generated by $a_1 T_R + a_2\sigma_3$.

Looking for constant spinor solutions of (\ref{eq:variationsN=2psi2}) and (\ref{eq:variationsN=2chi2}) we can apply the exterior covariant derivative to $\delta\psi_{\mu}$
thus turning the Killing spinor equation into an equation for the curvatures. Substituting the background (\ref{eq:N2Rsymmback}) we find 
\begin{eqnarray}
 \label{eq:N=2twistedkilling}
 2\partial_{[\mu}\delta\psi_{\nu]}^I & = & \left[\frac{1}{2}R_{\mu\nu}(\omega^{23})\g_{23} - R_{\mu\nu}(A)\mathi\g_5\right]\varepsilon^I
                                              - R_{\mu\nu}(V^3)(\mathi\sigma_3)^I_J\varepsilon^J \, = \nonumber\\
                                    & = & \mathi\Omega_{\mu\nu}\left[-\frac{\kappa}{2}\mathi\g_{23}\delta^I_J+a_1\g_5\delta^I_J+a_2(\sigma_3)^I_J\right] \varepsilon^J =0  \\
 \label{eq:N=2varchi2}
 \delta\chi^I & =& \frac{1}{2}\left[D - \frac{\kappa}{6}|e|\Omega_{23}\right]\varepsilon^I=0                   
 \end{eqnarray}
where (\ref{eq:N=2varchi2}) is obtained by substituting (\ref{eq:N=2twistedkilling}) in (\ref{eq:variationsN=2chi2})
and therefore it is only valid on the components of $\epsilon^I$ that are actual solutions of the Killing spinor equation.  

The $\chi^I$ variation can be set to zero by fixing the auxiliary field as $D=\frac{\kappa}{6}|e|\Omega_{23}$.
We are then left with a single defining equation for Killing spinors. 

%
%
%
%
%
%
\tocless\subsection{Twisting with flavors}\label{N=2 k=0}
%
%
%
%
%
Before solving the Killing spinor equation (\ref{eq:N=2twistedkilling}) we generalize the discussion to the case of 4d SCFTs admitting some global abelian non--R 
symmetry $U (1)_{flavor}$. Weakly gauging this symmetry implies turning on a non--vanishing background $\mathcal{N}= 2$ vector multiplet
$(B_\mu, X, \lambda^I, Y^A)$.
Such a multiplet contains one gauge field $B_\mu$ with curvature $R_{\mu\nu}(B)$, one complex
scalar $X$, two gaugini $\lambda^I$ forming a $SU (2)$ doublet, and one
auxiliary field $Y^A$ transforming in the adjoint of the R--symmetry group.
Setting the fermions $\lambda^I=0$, the supersymmetry variations of the bosonic components of the multiplet are identically vanishing, and they can be chosen to satisfy
\begin{equation}
 \label{eq:N=2gaugino1}
  \delta \lambda^I = \left[ \frac{1}{4} R_{ab}(B) \g^{ab}\delta^I_J + Y^A (\mathi\sigma_A)^I_J \right] \varepsilon^J \equiv 0 
\end{equation}
Gauging the global symmetry along $\Sigma$ with $R_{\mu\nu}(B)=b\,\Omega_{\mu\nu}$, and setting for instance $Y^{1,2}=0$, $Y^3=-\frac{b}{2}$ for the positive chirality component of $\varepsilon^J$ we obtain
\begin{equation}
 \label{eq:N=2gaugino2}
  \frac{b}{2}\left[\g^{23}\delta^I_J - (\mathi\sigma_3)^I_J\right] \epsilon^J = 0 \quad \Rightarrow \quad \left\{
   \begin{array}{l}
     (\g_{01}+1) \epsilon^1 = 0\\
     (\g_{01}-1) \epsilon^2 = 0 
   \end{array} \right.
\end{equation}
where we have used $\mathi\g_{23} = \g_{01} \g_5$ and $\g_5 \epsilon^J = \epsilon^J$. 

The previous condition is equivalent to requiring that the two components of the $\epsilon^I$ doublet have opposite chirality. Setting $Y^3=\frac{b}{2}$ would simply interchange the conditions on $\epsilon^1$ and $\epsilon^2$.

Another possibility to perform the flavor twist would be via a two--step procedure. We first gauge a $\mathcal{N}=1$ vector multiplet that breaks
explicitly $\mathcal{N}=2$ supersymmetry even before coupling the theory to a curved background. We then identify the $\mathcal{N}=1$ subsector
of the $\mathcal{N}=2$ theory which is compatible with this gauging, and apply the twist as in section \ref{sec:N=1}.
Observe that we could engineer such a reduction also in the absence of flavor symmetries.
In that case we should first perform a R--symmetry twist that preserves four 
supercharges. This twist would break R--symmetry and leave an unbroken $U(1)$ that could be treated as flavor symmetry useful for further twisting.
\\
%
%
%
%
%
%
\tocless\subsection{Classification of the solutions}\label{N=2 solutions}
%
%
%
%
%
In order to find solutions to eq. (\ref{eq:N=2twistedkilling}) we observe that the selected background breaks $Spin(3,1)\times SU(2)_R \rightarrow Spin(1,1)\times Spin(2)_\Sigma \times U(1)_{\sigma_3}$, and correspondingly the positive chirality components $\epsilon_\alpha^I$ in the $(\tmmathbf{2},\tmmathbf{2})$ representation as
\begin{equation}
 \label{eq:N=2representationsplit}
 \epsilon^I_\alpha \rightarrow \epsilon^1_+\oplus\epsilon^1_-\oplus\epsilon^2_+\oplus\epsilon^2_- 
\end{equation}
where on the r.h.s. $\pm$ indices denote the 2d chirality of the reduced spinors
\begin{equation}
 \label{eq:N=2chirality2d}
 \g_{01}\epsilon^I_\pm=\pm\epsilon^I_\pm\,  \qquad , \qquad \mathi\g_{23}\epsilon^I_\pm=\pm \epsilon^I_\pm 
\end{equation}
We can find solutions to (\ref{eq:N=2twistedkilling}) by appropriately choosing the values of the
twisting parameters $a_i$ as summarized in Table \ref{tab:N=2}.
\begin{table}[H]
 \centering
 \begin{tabular}{|c|c|}
  \hline
 supersymmetry  & $\delta\psi^I_\mu=0$\\
  \hline
  $\epsilon^1_+$ & $a_1 + a_2 - \kappa/2 = 0$\\
  $\epsilon^1_-$ & $a_1 + a_2 + \kappa/2 = 0$\\
  $\epsilon^2_+$ & $a_1 - a_2 - \kappa/2 = 0$\\
  $\epsilon^2_-$ & $a_1 - a_2 + \kappa/2 = 0$\\
  \hline
 \end{tabular}
 \caption{Supersymmetry equations for $\mathcal{N}=2$ theories. The supersymmetries in the left column
are preserved when the twisting parameters $a_i$ satisfy the corresponding equations in the column on the right.}
 \label{tab:N=2}
\end{table}
A further constraint comes from eq. (\ref{eq:N=2gaugino2}) when a global non--R symmetry is also gauged. 

\noindent
We discuss in detail the solutions for $\kappa \neq 0$ and $\kappa=0$, separately.

\vskip 10pt
\noindent
\underline{$\kappa \neq 0$}. For the case of non--zero curvature, we give a prototype of twist for each fixed amount of supersymmetry preserved in 2d. All the other choices are related by a trivial change of basis of the symmetries or a different choice of sign for the auxiliary fields.

\noindent
$\bullet$  For $a_1=-\frac{\kappa}{2}$ and $a_2 = 0$
  the preserved Killing spinors are $\epsilon^1_- \oplus \epsilon^2_-$ which form a $SU(2)_R$ doublet.
  The 4d R--symmetry is left unbroken and the 2d theory is a chiral $\mathcal{N}=(0,4)$ theory.
  If we add a flux for an external vector $B_\mu$, then equations (\ref{eq:N=2gaugino2})
  imply that only one of the two components of the doublet can be preserved according to the particular  
  choice of the auxiliary field $Y^A$ in the vector multiplet, hence supersymmetry is
  necessarily broken to $\mathcal{N}=(0,2)$.
  
\noindent
$\bullet$  For $a_1 = 0$ and $a_2 =-\frac{\kappa}{2}$
  the preserved supersymmetries are $\epsilon^1_- \oplus \epsilon^2_+$. R--symmetry is broken to $U(1)^2$ with generators
  $T_\pm\equiv\frac{1}{2}T_R \pm M_{23}$ and the preserved supersymmetry in two dimensions is  
  $\mathcal{N}=(2,2)$. The global symmetry generated by the background along
  $T\equiv M_{23}+\frac{1}{2}\sigma_3$ becomes a flavor symmetry in two dimensions since, by definition, the preserved supercharges transform trivially under it. In this case, gauging a global non--R symmetry with the corresponding connection $B_\mu$ together with the choice of auxiliary $Y^3 = - \tfrac{b}{2}$, does not constrain the Killing spinors any further (see eq. (\ref{eq:N=2gaugino2})) and the 2d theory maintains $\mathcal{N}=(2,2)$ supersymmetry. 
  
\noindent
$\bullet$ For $a_1 + a_2 = -\frac{\kappa}{2}$ the only preserved
  supersymmetry is $\epsilon^1_-$, hence the theory is $\mathcal{N}=(0,2)$ with $U(1)$ R--symmetry.
  In this case there are two new abelian flavor symmetries that were not present in the original
  4d theory, generated by the two combinations
  \begin{equation}
    T_1\equiv \frac{1}{2} T_R + M_{23}\quad\text{and}\quad
    T_2\equiv\frac{1}{2}\left(T_R - \sigma_3\right)
  \end{equation}
Turning on a flavor flux $B_\mu$ does not constrain this solution any further.

\vskip 12pt
\noindent 
\underline{$\kappa = 0$}. In the case of compactification on a torus we have two possible solutions.

\noindent
$\bullet$ The trivial solution corresponds to $a_1=a_2=0$, and $D=0$ in (\ref{eq:N=2varchi2}). This is the case where there is no twist, since 
the dimensional reduction on flat space preserves all supersymmetry.
The compactified theory flows to $\mathcal{N}= (4, 4)$ in 2d with global 
symmetry $SU (2) \times U (1)^2$ where the two abelian groups are 
generated by the combinations $T_\pm\equiv\frac{1}{2} T_R
\pm M_{23}$. Both sectors $(4, 0)$ and $(0, 4)$ provide a four dimensional
real representation of the $SU (2)$ R--symmetry group.

\noindent
$\bullet$  Another possible choice of supersymmetry preserving background on the torus corresponds to $a_1+a_2=0$ with both fluxes different from zero.
Solutions of (\ref{eq:N=2twistedkilling}) are then spinors $\epsilon^1_+\oplus\epsilon^1_-$ 
that transform trivially with respect to the 
background symmetry 
\begin{equation}
\label{eq:T}
T\equiv\frac{1}{2}\left(T_R - \sigma_3\right) 
\end{equation} 
The theory flows to $\mathcal{N}=(2,2)$ in 2d with $U(1)^2$ R--symmetry given by
\begin{equation}
 \label{eq:symmN=2k=0aneq0}
 T_\pm\equiv\frac{1}{2} T_R \pm M_{23}
\end{equation}
Turning on a background for an external global symmetry, $R_{\mu\nu}(B) = b\,\Omega_{\mu\nu}$, together with the auxiliary $Y^3 = - \tfrac{b}{2}$  further breaks supersymmetry to $\epsilon^1_-$,
as can be seen from (\ref{eq:N=2gaugino2}). In this case, the theory is $\mathcal{N}=(0,2)$ with $U(1)$
R--symmetry $T_R$ and two flavor symmetries which correspond precisely to the $T$ background (\ref{eq:T}) 
and the left R--symmetry $T_+$ (under which the right sector is invariant). Alternatively, choosing $Y^3 = + \tfrac{b}{2}$, the theory flows to $\mathcal{N}=(2,0)$ with two flavor symmetries $T$ and $T_-$.
\\
\\
The results of this section are summarized in the Table \ref{TableN2}.
\begin{table}[H]
\begin{center}
 \begin{tabular}{|c|cc|}   
  \hline
  $\kappa=0$ & $a_1 = a_2 = 0$ & $a_1 + a_2 = 0$ \\
  \hline
  $b   =  0$ & $\mathcal{N}=(4,4)$ & $\mathcal{N}=(2,2)$ \\
  $b \neq 0$ & $\mathcal{N}=(2,2)$ & $\mathcal{N}=(0,2)$ or  $(2,0)$ \\
  \hline
 \end{tabular}
\\~
\\
~
\\
 \begin{tabular}{|c|ccc|}   
  \hline
  $\kappa \neq 0$ & $a_1=-\frac{\kappa}{2}$, $a_2 = 0$ & $a_1=0$, $a_2 = -\frac{\kappa}{2}$ & $a_1+a_2=-\frac{\kappa}{2}$ \\
  \hline
  $b   =  0$ & $\mathcal{N}=(0,4)$ & $ \mathcal{N}=(2,2)$ & $\mathcal{N}=(0,2)$ \\
  $b \neq 0$ & $\mathcal{N}=(0,2)$ & $ \mathcal{N}=(2,2)$ & $\mathcal{N}=(0,2)$ \\
  \hline
 \end{tabular}
 \caption{Classification of topologically twisted 4d $\mathcal{N}=2$ SCFTs 
on Riemann surfaces of constant curvature $\kappa=\pm1,0$
in terms of the surviving amount of supersymmetry in 2d. We include 
the possibility of a twist along the flavor symmetries, with flux $b$. }
\label{TableN2}
\end{center}
\end{table}
%
%
%
%
%
\section{Twisted reduction of $\mathcal{N}=3$ SCFTs}
\label{sec:N=3}
%
%
%
%
%
%
It has been recently 
claimed \cite{Aharony:2015oyb,Garcia-Etxebarria:2015wns,Aharony:2016kai,Garcia-Etxebarria:2016erx}
 that 4d $\mathcal{N}=3$ SCFTs with no enhancement to $\mathcal{N}=4$ can exist at strong coupling. 
These theories have $SU(3)_R \times U(1)_R$ R--symmetry and their matter content coincides with the one of 4d $\mathcal{N}=4$ SYM. As a consequence there 
are no non--R global symmetries. 

Considering a $\mathcal{N}=3$ SCFT compactified on $M = \mathbbm{R}^{1,1} \times \Sigma$, a partial topological twist can be performed on $\Sigma$  using an abelian subgroup of the R--symmetry group.
In this section we study all possible solutions of the Killing spinor equations for such a twist, classifying all different configurations of preserved 
supercharges in two dimensions in terms of the different choices of the fluxes for the R--symmetry group.

As discussed above, the most natural framework where twisting a $\mathcal{N}=3$ SCFT on a curved manifold is $\mathcal{N}=3$ conformal supergravity 
\cite{Bergshoeff:1980is,Fradkin:1985am,vanNieuwenhuizen:1985dp}, whose Weyl multiplet and the corresponding non--linear supersymmetry transformations have been recently derived in \cite{vanMuiden:2017qsh}.
 
\begin{table}[H]
\centering
 \begin{tabular}{|c|ccccccc|cccc|}   
  \hline
  field     & $e^a_\mu$ & $b_\mu$ & $A_\mu$ & $V^A_\mu$ & $E_{I}$ & $T^{I}_{ab}$ & $D^{I}_{J}$ & $\psi^I_\mu$ & $\Lambda$ & $\chi_{IJ}$ & $\zeta^I$ \\
  \hline
  $SU(3)_R\times U(1)_R$ & $\tmmathbf{1}_0$ & $\tmmathbf{1}_0$ & $\tmmathbf{1}_0$ & $\tmmathbf{8}_0$ & $\tmmathbf{\bar{3}}_{2}$ & $\tmmathbf{3}_{-2}$ & $\tmmathbf{8}_0$ & $\tmmathbf{3}_1$ & $\tmmathbf{1}_{3}$ & $\tmmathbf{6}_1$ & $\tmmathbf{3}_1$ \\
  $\#$ of real d.o.f. & $5$ & $0$ & $3$ & $24$ & $6$ & $18$ & $8$ & $24$ & $4$ & $24$ & $12$\\
  \hline
 \end{tabular}
\caption{Field content of the Weyl multiplet in $\mathcal{N}=3$ conformal supergravity.}
\label{tab:N=3weyl}
\end{table}
The $\mathcal{N}=3$ Weyl multiplet in four dimensions is given in Table \ref{tab:N=3weyl}.
In particular, $A_\mu$ and $V_\mu^A$, $A = 1, \cdots , 8$ are the gauge fields associated to the R--symmetry $U(1)_R$ and $SU(3)_R$ transformations, respectively.

The R--symmetry group $SU(3)_R$ is generated by antihermitian matrices
$(\mathi\lambda_A)$, with $A=1,..,8$. We choose a basis in which the $SU(3)$ can be embedded into the
top left $3\times 3$ block of $SU(4)$, so that the first 8 generators of $SU(4)$ reduce
straightforwardly to the generators of $SU(3)$. 
The $U(1)_R$ group is obtained by mixing the $U(1)$ from the decomposition of $SU(4)_R$ into $SU(3)_R \times U(1)$
and the chiral $U(1)$ that enhances the superalgebra from $PSU(2,2|4)$ to $SU(2,2|4)$ \cite{Bergshoeff:1980is,Ferrara:1998zt}.
We observe that these two $U(1)$ groups act proportionally to each other on the components of the $\mathcal{N}=4$ Weyl multiplet that survive in the
projection to the $\mathcal{N}=3$ Weyl multiplet.

As in the previous cases, we are interested in preserving supersymmetry while coupling the SCFT to a
curved background describing the geometry of the manifold $M$. We choose a background Weyl multiplet where, together with the fermions, all the bosonic fields are set to zero except for $e^a_\mu$, $A_\mu$, $V_\mu^A$ and $D^{I}_{J}$.
Consequently, the conditions for the fermion variations to vanish read \cite{vanMuiden:2017qsh}
\begin{eqnarray}
 \label{eq:N=3varpsi}
 \delta\psi^I_\mu & = & \left[\partial_\mu + \frac{1}{4}\omega^{ab}_\mu\g_{ab}
                        -A_\mu\mathi\g_5 \right] \varepsilon^I
                        -V^A_\mu(\mathi\lambda_A)^I_J\varepsilon^J =0 \\
 \label{eq:N=3varchi}
 \delta\chi_{IJ} & = & -\frac{1}{2}\varepsilon_{KL(I} D^K_{J)}\varepsilon^L
                       -\frac{1}{4}\varepsilon_{KL(I}\g^{ab}R_{ab}(V^A)(\mathi\lambda_A)^{K}_{J)}\varepsilon^L =0  \\
 \label{eq:N=3varzeta}
 \delta\zeta^I & = & \frac{1}{4}D^{I}_{K}\varepsilon^K
                     -\frac{1}{24}\g^{ab}R_{ab}(V^A)(\mathi\lambda_A)^{I}_K\varepsilon^{K}
                     +\frac{1}{3}\g^{ab}R_{ab}(A)\mathi\g_5\varepsilon^{I} =0 \\
 \label{eq:N=3varLambda}
 \delta\Lambda & = & 0 
\end{eqnarray}
These provide the set of constraints that select the surviving Killing spinors in two dimensions.  
In order to find non--trivial solutions, we choose the R--symmetry $V_\mu^A$ and $A_\mu$ background fields such that 
\begin{equation}
 \label{eq:N=3twist2}
 R_{\mu\nu}(V^3) = -a_1 \Omega_{\mu\nu}~,\quad R_{\mu\nu}(V^8) = -\sqrt{3} a_2 \Omega_{\mu\nu}~,\quad R_{\mu\nu}(V^A) = 0\quad\text{for}\quad A\neq 3,8
\end{equation}
\begin{equation}
 \label{eq:N=3twist1}
 R_{\mu\nu}(A) = -a_3 \Omega_{\mu\nu}
\end{equation}
and subject to appropriate quantization conditions (see the remark at the end of section \ref{sec:N=4}).
The non--trivial Killing spinor equations then reduce to
\begin{eqnarray}
 \label{eq:killingN=3}
 2\partial_{[\mu}\delta\psi^I_{\nu]} & = & \frac{1}{2}R_{\mu\nu}(\omega^{23})\g_{23}\varepsilon^I
                                     - R_{\mu\nu}(A)\mathi\g_5\varepsilon^I
                                     - \left[ R_{\mu\nu}(V^3)(\mathi\lambda_3)^I_J +
                                              R_{\mu\nu}(V^8)(\mathi\lambda_8)^I_J \right]\varepsilon^J \nonumber\\
    & = & \mathi\Omega_{\mu\nu}\left[-\frac{\kappa}{2}\mathi\g_{23}\delta^I_J
          +a_1(\lambda_{3})^I_J
          +a_2\sqrt{3}(\lambda_{ 8})^I_J
          +a_3\g_5\delta^I_J\right]\varepsilon^J = 0 
\end{eqnarray}
 together with the two auxiliary conditions (\ref{eq:N=3varchi}, \ref{eq:N=3varzeta}).
%
%
%
%
%
%
\tocless\subsection{Classification of the solutions}\label{N=3 solutions}
%
%
%
%
%
In order to find non--trivial solutions to equation (\ref{eq:killingN=3}) we restrict
the discussion to the positive chirality components of the $\varepsilon^I$ spinors. We observe that under the breaking $Spin(3,1)\times SU(3)_R \times U(1)_R 
\rightarrow Spin(1,1)\times Spin(2)_\Sigma \times U(1)_{\lambda_3} \times U(1)_{\lambda_8} \times U(1)_R $ realized by the chosen geometry, the original 4d chiral parameters $\epsilon^I_\alpha$, $I=1,2,3$, split as 
\begin{equation}
\epsilon^I_\alpha \rightarrow \epsilon_+^1 \oplus  \epsilon_-^1 \oplus \epsilon_+^2 \oplus  \epsilon_-^2 \oplus \epsilon_+^3 \oplus  \epsilon_-^3 
\end{equation}
where $\pm$ still indicate the 2d chirality as defined in (\ref{eq:N=2chirality2d}). The spinors
are charged under $U(1)_{\lambda_3} \times U(1)_{\lambda_8} \times U(1)_R $ according to
$\epsilon^1_\pm \to (1,\frac{1}{\sqrt{3}},1)$, $\epsilon^2_\pm \to (-1,\frac{1}{\sqrt{3}},1)$ and $\epsilon^3_\pm \to (0,-\frac{2}{\sqrt{3}},1)$.
Supersymmetry preserving equations are then given in Table \ref{tab:N=3}.
\begin{table}[H]
 \centering
 \begin{tabular}{|c|r|}
  \hline
  supersymmetry    & $\delta\psi^I_\mu=0$ ~~~~~~~~~~ \\
  \hline
  $\epsilon^1_\pm$ & $  a_1 +   a_2 + a_3 \mp \kappa/2 = 0$ \\
  $\epsilon^2_\pm$ & $ -a_1 +   a_2 + a_3 \mp \kappa/2 = 0$ \\
  $\epsilon^3_\pm$ & $      - 2 a_2 + a_3 \mp \kappa/2 = 0$ \\
  \hline
 \end{tabular}
 \caption{Supersymmetry equations for $\mathcal{N}=3$ theories.}
 \label{tab:N=3}
\end{table}
Once the equation  $\delta\psi^I_\mu=0$ has been solved for a particular set of $a_i$ parameters, equations (\ref{eq:N=3varchi}, \ref{eq:N=3varzeta}) 
need to be satisfied. In Appendix \ref{app:B} we prove that solutions to $\delta\chi_{IJ}=0$ and
$\delta\zeta^{I}=0$ always exist if we appropriately choose the value of the components of the auxiliary field $D^I_J$.

In Table \ref{TableN3} we list all possible solutions to the conditions in Table \ref{tab:N=3} together with the corresponding preserved supersymmetries and the remaining 2d R--symmetry. We focus on the cases with mostly right supersymmetry and 
for each possibility we pick up just a choice of fluxes. All the other possibilities can be obtained through a change of basis for the $SU(3)_R$ generators.
\begin{table}[H]
\begin{center}
\begin{tabular} {|c|c|c|c|}
\hline
$\kappa = 0$ & fluxes & supersymmetries & R--symmetry \\
\hline
 \begin{tabular}{c}
    $\mathcal{N}=(6,6)$ \\
    $\mathcal{N}=(4,4)$ \\
    $\mathcal{N}=(2,2)$ \\
 \end{tabular}
 &
 \begin{tabular}{c}
  $a_1=0$, $a_2=0$, $a_3=0$\\
  $a_1=0$, $a_2+a_3=0$\\
  $a_1+a_2+a_3=0$\\
 \end{tabular}
 &
 \begin{tabular}{c}
  $\epsilon^1_\pm\oplus\epsilon^2_\pm\oplus\epsilon^3_\pm$\\
  $\epsilon^1_\pm\oplus\epsilon^2_\pm$\\
  $\epsilon^1_\pm$\\
 \end{tabular}
 &
 \begin{tabular}{c}
    $SU(3)\times U(1)$ \\
    $SU(2) \times U(1)$ \\
    $U(1)$ \\
 \end{tabular}
 \\
\hline
\end{tabular}

\vspace{0.2cm}

\begin{tabular} {|c|c|c|c|}
\hline
$\kappa \neq 0$ & fluxes & supersymmetries & R--symmetry \\
 \hline
 \begin{tabular}{c}
    $\mathcal{N}=(2,4)$ \\
    $\mathcal{N}=(0,6)$ \\
    $\mathcal{N}=(2,2)$ \\
    $\mathcal{N}=(0,4)$ \\
    $\mathcal{N}=(0,2)$ \\
 \end{tabular}
 &
 \begin{tabular}{c}
  $a_1=0$, $a_2=-\frac{\kappa}{3}$, $a_3=-\frac{\kappa}{6}$\\
  $a_1=0$, $a_2=0$, $a_3=-\frac{\kappa}{2}$\\
  $a_1=-\frac{\kappa}{2}$, $a_2+a_3=0$\\
  $a_1=0$, $a_2+a_3=-\frac{\kappa}{2}$\\
  $a_1+a_2+a_3=-\frac{\kappa}{2}$\\
 \end{tabular}
 &
 \begin{tabular}{c}
  $\epsilon^3_+\oplus\epsilon^1_-\oplus\epsilon^2_-$\\
  $\epsilon^1_-\oplus\epsilon^2_-\oplus\epsilon^3_-$\\
  $\epsilon^2_+\oplus\epsilon^1_-$\\
  $\epsilon^1_-\oplus\epsilon^2_-$\\
  $\epsilon^1_-$\\
 \end{tabular}
 &
 \begin{tabular}{c}
    $SU(2) \times U(1)$ \\
    $SU(3) \times U(1)$ \\
    $U(1) \times U(1)$ \\
    $SU(2) \times U(1)$ \\
    $U(1)$ \\
 \end{tabular}
 \\
 \hline
\end{tabular}
 \caption{Classification of topologically twisted 4d $\mathcal{N}=3$ SCFTs 
on  constant curvature Riemann surfaces 
in terms of the surviving amount of supersymmetry in 2d. 
In the last column we indicate the subgroup of 4d R--symmetry that is compatible with the twisted compactification.}
\label{TableN3}
\end{center}
\end{table}
In all the $\kappa \neq 0$ cases a $U(1)$ flavor symmetry survives in two dimensions, being it associated to the diagonal generator $(\frac{\kappa}{2} \mathi \gamma_{23} - T)$, where $T= a_1 \lambda_3 + a_2 \sqrt{3} \lambda_8 + a_3  \g_{5}$, under which, by definition, the surviving Killing spinors are neutral. However, in the $\mathcal{N}=(2,4)$ case, one extra $U(1)$  symmetry emerges from the topological twist, which is generated  by $T$ itself (or any linear combination of $T$ with the flavor symmetry generator).
Although under $T$ the supercharges are charged, this symmetry cannot be a R-symmetry of the low energy SCFT. It might be that this symmetry is not a symmetry of the 2d theory, or appears as an outer automorphism of the 2d supersymmetry algebra \footnote{We are grateful to Nikolay Bobev for raising this interesting point.}. However, in order to get more insight on it one should know the actual SCFT algebra that emerges from the twisted reduction and the relation of $T$ with the rest of the superalgebra generators.

From Table \ref{TableN3}  we note that, while for $\kappa \neq 0$ we can reduce supersymmetry in two dimensions to 
 $\mathcal{N}=(0,2)$, in the case of the torus the minimum amount of supersymmetry that we obtain by partial topological twist is $\mathcal{N}=(2,2)$. This is a consequence of the fact that in the ${\mathcal N}=3$ case there are no flavor symmetries that can be weakly gauged in order to further reduce supersymmetry. 
 
However, also in the $\kappa=0$ case we can reduce supersymmetry to $\mathcal{N}=(0,2)$ 
 by a two--step procedure similar to the one already discussed in section \ref{sec:N=2}
for $\mathcal{N}=2$ theories without flavor symmetries.
This works as follows.
First we perform a R--symmetry twist that preserves either four or eight supercharges.
This twist breaks R--symmetry as well, leaving some flavor symmetries with the associated vector multiplets. 
The second step of this reduction is performed by introducing a ($\mathcal{N}=1$ or $\mathcal{N}=2$) background for the
vector multiplet that preserves only half of the supercharges.
For example, if we use this procedure in the case of $a_1+a_2+a_3=0$
we preserve in the first step a 4d $\mathcal{N}=1$ subalgebra of the original 
$\mathcal{N}=3$. The leftover R--symmetry is just $U(1)$, while 
the residual $SU(2)\times U(1)$ from the original $SU(3)_R \times U(1)_R$ 
survives as flavor symmetry. In the second step we can gauge an abelian subgroup of this
flavor symmetry. The corresponding gaugino background then breaks supersymmetry to
$\mathcal{N}=(2,0)$ or $\mathcal{N}=(0,2)$ as we can see from (\ref{eq:gaugino-variation}).
\\

%
%
%
%
%
%
\section{Twisted reduction of $\mathcal{N}=4$ SCFTs} 
\label{sec:N=4}
%
%
%
%
%
%
%
%
This case has been extensively discussed in the literature \cite{Bershadsky:1995vm,Bershadsky:1995qy,Maldacena:2000mw,
Benini:2012cz,Benini:2013cda}. For completeness, here we briefly review the main results in the language of conformal supergravity.

The supercharges are in the antifundamental representation of the $SU(4)_R$ R--symmetry group 
The generators are
traceless hermitian matrices $\lambda_A$, $A=1,...,15$. We choose a basis in which the Cartan
subalgebra is spanned by
\begin{equation}
 \label{eq:SU4algebra}
 \lambda_3 = \left[
    \begin{array}{cccc}
        1 & 0 & 0 & 0 \\
        0 & -1 & 0 & 0 \\
        0 & 0 & 0 & 0 \\
        0 & 0 & 0 & 0
    \end{array}
 \right]
 \quad
 \lambda_8 = \frac{1}{\sqrt{3}}\left[
    \begin{array}{cccc}
        1 & 0 & 0 & 0 \\
        0 & 1 & 0 & 0 \\
        0 & 0 & -2 & 0 \\
        0 & 0 & 0 & 0
    \end{array}
 \right]
 \quad
 \lambda_{15} = \frac{1}{\sqrt{6}}\left[
    \begin{array}{cccc}
        1 & 0 & 0 & 0 \\
        0 & 1 & 0 & 0 \\
        0 & 0 & 1 & 0 \\
        0 & 0 & 0 & -3
    \end{array}
 \right]
\end{equation}

The Weyl multiplet of the $\mathcal{N}=4$ conformal supergravity contains the gauge fields $e^a_\mu$, $b_\mu$
$V^A_\mu$ and $\psi^I_\mu$, the bosonic auxiliary fields $C$, $E_{IJ}$, $T^{IJ}_{ab}$ $D^{IJ}_{KL}$
and the fermionic auxiliaries $\Lambda_I$, $\chi^{IJ}_K$. In Table \ref{tab:N=4weyl} we list the corresponding $SU(4)_R$ representations. 
For a complete description of $\mathcal{N}=4$ supergravity we refer to \cite{Bergshoeff:1980is, Fradkin:1985am}.
\begin{table}[H]
 \centering
 \begin{tabular}{|c|ccccccc|ccc|}   
  \hline
  field     & $e^a_\mu$ & $b_\mu$ & $V^A_\mu$ & $C$ & $E_{IJ}$ & $T^{IJ}_{ab}$ & $D^{IJ}_{KL}$ & $\psi^I_\mu$ & $\Lambda_I$ & $\chi^{IJ}_K$ \\
  \hline
  $SU(4)_R$ & $\tmmathbf{1}$ & $\tmmathbf{1}$ & $\tmmathbf{15}$ & $\tmmathbf{1}$ & $\tmmathbf{\bar{10}}$ & $\tmmathbf{6}$ & $\tmmathbf{20}$ & $\tmmathbf{4}$ & $\tmmathbf{\bar{4}}$ & $\tmmathbf{20}$ \\
  $\#$ of real d.o.f. & $5$ & $0$ & $45$ & $2$ & $20$ & $36$ & $20$ & $32$ & $16$ & $80$\\
  \hline
 \end{tabular}
\caption{Field content of the Weyl multiplet in $\mathcal{N}=4$ conformal supergravity.}
\label{tab:N=4weyl}
\end{table}
As in the previous cases, we define the theory on the curved manifold\footnote{Four dimensional ${\cal N}=4$ superconformal theories on curved backgrounds have been considered in \cite{Maxfield:2016lok}.} $M=\mathbbm{R}^{1,1} \times \Sigma$,
by freezing the Weyl multiplet to contain as only non--vanishing components the vielbein, a R--symmetry background $V_\mu^A$ and an auxiliary 
field $D^{IJ}_{KL}$. 
Supersymmetry is (partially) preserved if there exist spinor parameters $\varepsilon^I_\alpha$ satisfying
\begin{eqnarray}
 \label{eq:N=4varpsi}
 \delta\psi^I_\mu & = & \partial_\mu\varepsilon^I+\frac{1}{4}\omega^{ab}_\mu\g_{ab}\varepsilon^I- V^A_\mu(\mathi\lambda_A)^I_J\varepsilon^J =0 \\
 \label{eq:N=4varchi}
 \delta\chi^{IJ}_K & = & \frac{1}{2}D^{IJ}_{KL}\varepsilon^L-\frac{1}{2}\g^{ab}R_{ab}(V^A)(\mathi\lambda_A)^{[I}_K\varepsilon^{J]}
                         -\frac{1}{6}\g^{ab}\delta^{[I}_K R_{ab}(V^A)(\mathi\lambda_A)^{J]}_L \varepsilon^L =0
                         \end{eqnarray}
while $\delta\Lambda_I$ is identically zero in the selected background. 
In order to find non--trivial solutions we choose the R--symmetry gauge field such that 
\begin{equation}
 \label{eq:N=4twist2}
 R_{\mu\nu}(V^3) = -a_1 \Omega_{\mu\nu},\quad R_{\mu\nu}(V^8) = -\sqrt{3} a_2 \Omega_{\mu\nu},\quad R_{\mu\nu}(V^{15}) = -\sqrt{6} a_3 \Omega_{\mu\nu},
\end{equation}
\begin{equation}
 \label{eq:N=4twist1}
 R_{\mu\nu}(V^A) = 0\quad\text{for}\quad A\neq 3,8,15
\end{equation}
subject to appropriate quantization conditions (see the remark at the end of this section).
Equations (\ref{eq:N=4varpsi}) and (\ref{eq:N=4varchi}) then reduce to
\begin{eqnarray}
 \label{eq:killingN=4}
 2\partial_{[\mu}\delta\psi^I_{\nu]} & = & \frac{1}{2}R_{\mu\nu}(\omega^{23})\g_{23}\varepsilon^I
                                     - R_{\mu\nu}(V^A)(\mathi\lambda_A)^I_J \varepsilon^J \nonumber\\
    & = & \mathi\Omega_{\mu\nu}\left[-\frac{\kappa}{2}\mathi\g_{23}\delta^I_J
          +a_1(\lambda_{3})^I_J
          +a_2\sqrt{3}(\lambda_{ 8})^I_J
          +a_3\sqrt{6}(\lambda_{15})^I_J\right]\varepsilon^J = 0 
\end{eqnarray}

%
%
%
%
%
%
\tocless\subsection{Classification of the solutions}\label{N=4 solutions}
%
%
%
%
%
The selected background induces the breaking $Spin(3,1)\times SU(4)_R \
\rightarrow Spin(1,1)\times Spin(2)_\Sigma \times U(1)_{\lambda_3} \times U(1)_{\lambda_8} \times U(1)_{\lambda_{15}}$
under which the chiral supersymmetry parameters split as
\begin{equation}
\epsilon^I_\alpha \rightarrow \epsilon_+^1 \oplus  \epsilon_-^1 \oplus \epsilon_+^2 \oplus  \epsilon_-^2 \oplus \epsilon_+^3 \oplus  \epsilon_-^3\oplus \epsilon_+^4 \oplus  \epsilon_-^4
\end{equation}
where, once again, the $\pm$ indices indicate chirality as defined in (\ref{eq:N=2chirality2d}). The spinors are charged under $U(1)_{\lambda_3} \times U(1)_{\lambda_8} \times U(1)_{\lambda_{15}}$ according to $\epsilon^1_\pm \to (1,\frac{1}{\sqrt{3}},\frac{1}{\sqrt{6}})$, $\epsilon^2_\pm \to (-1,\frac{1}{\sqrt{3}},\frac{1}{\sqrt{6}})$, $\epsilon^3_\pm \to (0,-\frac{2}{\sqrt{3}},\frac{1}{\sqrt{6}})$ and $\epsilon^4_\pm \to (0,0,-\frac{3}{\sqrt{6}})$.

Therefore, equation (\ref{eq:killingN=4}) translates into the set of supersymmetry preserving equations listed in Table \ref{tab:N=4}.
\begin{table}[H]
 \centering
 \begin{tabular}{|c|r|}
  \hline
  supersymmetry    & $\delta\psi^I_\mu=0$ ~~~~~~~~~~ \\
  \hline
  $\epsilon^1_\pm$ & $ a_1 +   a_2 +   a_3 \mp \kappa/2 = 0$ \\
  $\epsilon^2_\pm$ & $-a_1 +   a_2 +   a_3 \mp \kappa/2 = 0$ \\
  $\epsilon^3_\pm$ & $     - 2 a_2 +   a_3 \mp \kappa/2 = 0$ \\
  $\epsilon^4_\pm$ & $             - 3 a_3 \mp \kappa/2 = 0$ \\
  \hline
 \end{tabular}
 \caption{Supersymmetry equations for $\mathcal{N}=4$ theories.}
 \label{tab:N=4}
\end{table}
For any set of $a_i$ parameters satisfying one of the conditions in the previous table, equation (\ref{eq:N=4varchi}) can be satisfied
by a suitable choice of the background auxiliary fields $D^{IJ}_{KL}$ without further constraining the $\epsilon^I$ parameters.

In Table \ref{TableN4} we list explicit solutions for the $a_i$ parameters and the corresponding 2d surviving supersymmetry with its R--symmetry group. 
We focus on the cases with mostly right--handed supersymmetry and for each possibility we pick up just one particular configuration of fluxes.
\begin{table}[H]
\begin{center}
\begin{tabular} {|c|c|c|c|}
\hline
$\kappa = 0$ & fluxes & supersymmetries & R--symmetry \\
\hline
 \begin{tabular}{c}
    $\mathcal{N}=(8,8)$ \\
    $\mathcal{N}=(4,4)$ \\
    $\mathcal{N}=(2,2)$ \\
 \end{tabular}
 &
 \begin{tabular}{c}
  $a_1=0$, $a_2=0$, $a_3=0$\\
  $a_1=0$, $a_2+a_3=0$\\
  $a_1+a_2+a_3=0$\\
 \end{tabular}
 &
 \begin{tabular}{c}
  $\epsilon^1_\pm\oplus\epsilon^2_\pm\oplus\epsilon^3_\pm\oplus\epsilon^4_\pm$\\
  $\epsilon^1_\pm\oplus\epsilon^2_\pm$\\
  $\epsilon^1_\pm$\\
 \end{tabular}
 &
 \begin{tabular}{c}
    $SU(4)$ \\
    $SU(2) \times U(1)$ \\
    $U(1)$ \\
 \end{tabular}
 \\
\hline
\end{tabular}

\vspace{0.2cm}

\begin{tabular} {|c|c|c|c|}
\hline
$\kappa \neq 0$ & fluxes & supersymmetries & R--symmetry \\
 \hline
 \begin{tabular}{c}
    $\mathcal{N}=(4,4)$ \\
    $\mathcal{N}=(0,6)$ \\
    $\mathcal{N}=(2,2)$ \\
    $\mathcal{N}=(0,4)$ \\
    $\mathcal{N}=(0,2)$ \\
 \end{tabular}
 &
 \begin{tabular}{c}
  $a_1=0$, $a_2=-\frac{\kappa}{3}$, $a_3=-\frac{\kappa}{6}$\\
  $a_1=0$, $a_2=0$, $a_3=-\frac{\kappa}{2}$\\
  $a_1=-\frac{\kappa}{2}$, $a_2+a_3=0$\\
  $a_1=0$, $a_2+a_3=-\frac{\kappa}{2}$\\
  $a_1+a_2+a_3=-\frac{\kappa}{2}$\\
 \end{tabular}
 &
 \begin{tabular}{c}
  $\epsilon^3_+\oplus\epsilon^4_+\oplus\epsilon^1_-\oplus\epsilon^2_-$\\
  $\epsilon^1_-\oplus\epsilon^2_-\oplus\epsilon^3_-$\\
  $\epsilon^2_+\oplus\epsilon^1_-$\\
  $\epsilon^1_-\oplus\epsilon^2_-$\\
  $\epsilon^1_-$\\
 \end{tabular}
 &
 \begin{tabular}{c}
    $SU(2) \times SU(2)
    $ \\
    $SU(3) \times U(1)$ \\
    $U(1) \times U(1)$ \\
    $SU(2) \times U(1)$ \\
    $U(1)$ \\
 \end{tabular}
 \\
 \hline
\end{tabular}
\caption{Classification of topologically twisted 4d $\mathcal{N}=4$ SYM on Riemann surfaces of constant curvature $\kappa=\pm1,0$ in terms of the surviving amount of supersymmetry in 2d. In the last column we indicate the subgroup of 4d R--symmetry that is compatible with the twisted compactification.}
\label{TableN4}
\end{center}
\end{table}
Similarly to what happens in the ${\cal N}=3$ case, for the $\mathcal{N}=(4,4)$ solution with $\kappa\neq0$ one extra $U(1)$ symmetry generated by $T = a_1 \lambda_3 + a_2 \sqrt{3} \lambda_8 + a_3 \sqrt{6} \lambda_{15}$ emerges from the topological twist. Although $T$ acts non--trivially on the supercharges, this cannot be a R-symmetry of the low energy SCFT, but it could be identified as an outer automorphism of the 2d superconformal algebra.

We conclude this analysis by observing that, as in the case of $\mathcal{N}=3$ theories, although there are no flavor symmetries,
we can further reduce supersymmetry by performing a two step reduction. 
The first step consists of turning on an R--symmetry twist, breaking supersymmetry to $\mathcal{N}=2$ or $\mathcal{N}=1$.
The second step consists of introducing a background $\mathcal{N}=2$ or $\mathcal{N}=1$ vector multiplet for the leftover non--R 
\emph{flavor} symmetry, such that only half of the supercharges are preserved.\\

\noindent
{\bf Remark:} In the ${\cal N}=3,4$ cases the background quantization conditions $a_i \nu \in {\mathbbm{Z}}$ used for ${\cal N}=1,2$ are too restrictive, but fortunately they can be partially relaxed. For example, if we look at the $\mathcal{N}=(4,4)$, $\kappa\neq0$ case in Table \ref{TableN4} the solutions $a_2=-\kappa/3$ and $a_3=-\kappa/6$ would be incompatible with such a quantization condition
and consequently the R--symmetry bundle would be ill--defined. However, in this case the quantization condition that one has to 
actually impose is that the combination $T\equiv a_2\sqrt{3}\lambda_8+a_3\sqrt{6}\lambda_{15}$
(i.e., the background symmetry that has been gauged by the twist) assigns integer charges to every
field/representation of the theory. Substituting the explicit values of $a_2$ and $a_3$ we can see that
the background symmetry $T$ corresponds precisely to the $U(1)$ R--symmetry of the $\mathcal{N}=(4,4)$
theory
\begin{equation}
 T = -\frac{\kappa}{2} \left[\frac{2}{3}(\sqrt{3}\lambda_8)+\frac{1}{3}(\sqrt{6}\lambda_{15})\right]=-\frac{\kappa}{2}
     \left[
      \begin{array}{cccc}
        1 & 0 & 0 & 0 \\
        0 & 1 & 0 & 0 \\
        0 & 0 & -1 & 0 \\
        0 & 0 & 0 & -1
      \end{array}
     \right]
\end{equation} 
The quantization condition then becomes $\frac{\kappa}{2}\nu\in\mathbbm{Z}$, which is satisfied for any choice of genus $g$. A similar analysis applies to the other cases, leading to the same conclusion.
\\

\section{c-extremization for 2d $ \mathcal{N}=(0,2)$ SCFTs} \label{sec:central}

In this section we focus on the special case of two dimensional $\mathcal{N}=(0,2)$ theories obtained by twisted compactification of $\mathcal{N}$--extended supersymmetric theories in four dimensions, as described in the previous sections. In particular, we determine a general expression for the central charge and the other 2d global anomalies.

Generalizing the prescription developed in \cite{Benini:2015bwz} for $\mathcal{N}=1$ SCFT's, we begin with the 4d anomaly polynomial $I_6$ for the $U(1)$ global symmetries, including the abelian symmetry coupled to the twisting supergravity background, and integrate it along the $\Sigma$ directions. The resulting expression is a 4--form that can be identified with the anomaly polynomial $I_4$ of the 2d theory.  
From this expression we can then infer the 2d anomalies as functions of the 4d anomalies and of the background fluxes. 

In this procedure we have to take into account that, even if the R--symmetry we start with is the exact R--symmetry in 4d, along the dimensional flow the $U(1)_R$ can mix with other abelian flavor symmetries. The exact 2d central charge is then reconstructed by extremizing a trial central charge as a function of the mixing coefficients \cite{Benini:2012cz}. Because of this potential mixing, in the reduction procedure we can start with any trial $U(1)$ R--symmetry $T_R$ in four dimensions, as different choices will simply shift the mixing parameters of the 2d theory without affecting the final result of the extremization procedure.

We consider a generic SCFT in four dimensions with different amount of supersymmetry that flows to a
$\mathcal{N}=(0,2)$ theory in two dimensions. As turns out to be clear from our discussion in section \ref{sec:N=1}, 
in the $\mathcal{N}=1$ case the 4d trial $T_R$ generator can be identified with the original $U(1)$ R--symmetry generator of the $\mathcal{N}=1$ algebra.
Calling $t_R$ the corresponding abelian generator in the reduced $\mathcal{N}=(0,2)$
theory, in general the two $U(1)$ symmetries will have different matrix forms, but they can be identified up to a mixing with the
abelian flavor symmetries
\begin{equation}
 \label{eq:mixR-symmFlav}
 T_R \rightarrow t_R + \sum_{i=1}^n \xi_i t_i
\end{equation}
where $t_i$ are the generators of the abelian flavor symmetries $U(1)_i$ in the 2d representation,
while $\xi_i$ are the mixing coefficients.
The relation (\ref{eq:mixR-symmFlav}) represents the most general trial 2d R--current, 
involving abelian currents that do not necessarily mix with the R--current in the 4d SCFT, 
as the baryonic symmetries in toric quiver gauge theories \cite{Bertolini:2004xf,Butti:2005vn}.

Our discussion can be applied also to the case of extended supersymmetry.
In that case we can identify the generator $t_R$ with the four dimensional
R--current of the $\mathcal{N}=1$ subalgebra. 
When reducing to 2d $\mathcal{N}=(0,2)$ all the other abelian global currents
have to be treated as flavor symmetries that can potentially mix with the 2d $R$--symmetry.
In the rest of this section we restrict to the case of 4d $\mathcal{N}=1$ SCFT.

In order to compute the anomaly polynomial $I_6$, which encodes all the global and gravitational
anomalies of the twisted theory\footnote{The gauge theory is assumed to be free of local gauge
anomalies, i.e., anomalies for symmetries coupled to dynamical gauge vectors.}, we first couple
each global symmetry to a background connection on $\mathbbm{R}^{1,1}$, which being topologically
trivial can be compactified into a torus $\mathbbm{T}\cong S^1\times S^1$. The topological twist
introduces additional background components for $U(1)_R$ and $U(1)_i$ also along the $\Sigma$ directions.

Following the notations of Appendix \ref{app:A}, if we denote $f_R$ the first Chern class of the
R--symmetry bundle and $f_i$ the class associated to the gauging of the abelian $U(1)_i$ flavor symmetries,
then we can write
\begin{equation}
 \label{eq:twistchernclass}
 f_R = f_R^\mathbbm{T} + f_R^\Sigma \qquad \text{and} \qquad f_i = f_i^\mathbbm{T} + f_i^\Sigma
\end{equation}
where, the components in the direction of $\Sigma$ are defined by (\ref{eq:curvatureN=1}) and (\ref{eq:flavorN=1gauging}) as 
\begin{equation}
 \label{eq:twistchernclass2}
 f_R^\Sigma = -a\left[\frac{\Omega}{2\pi}\right] \qquad \text{and} \qquad f_i^\Sigma = b_i\left[\frac{\Omega}{2\pi}\right]
\end{equation}
so that the total Chern class of the global symmetry bundle $E$ (see Appendix \ref{app:A} for the definition)
restricted to the Riemann surface $\Sigma$ is
\begin{equation}
 \label{eq:twistchernclass3}
 c_1(E)\Big|_\Sigma = \tmop{Tr}[T_R] f_R^\Sigma + \sum_i \tmop{Tr}[T_i] f_i^\Sigma = \tmop{Tr} [-a T_R + {\textstyle\sum_i} b_i T_i]\left[\frac{\Omega}{2\pi}\right]
\end{equation}
where $T_R$ and $T_i$ are the 4d generators and the trace means summing over positive (negative) chirality fermions with plus (minus) sign. 
Here the twisting parameter $a$ is fixed by the Killing spinor equation (\ref{eq:twistedKillingSpinorEq3})
to the value $-\frac{\kappa}{2}$. We can then interpret the combination $T\equiv \frac{\kappa}{2}T_R+\sum_i b_i T_i$
to be the abelian symmetry which generates the topological twist on $\Sigma$.

According to formula (\ref{anomalypol}), the anomaly polynomial is given by the six--form
\begin{eqnarray}
  I_6 & = &   \tmop{ch}_3 (E) - \frac{1}{24} p_1 (M) \tmop{ch}_1 (E) \nonumber\\
      & = &   \frac{1}{6} \tmop{Tr} [T_R^3] f_R^3 + \frac{1}{2} \sum_{i} \tmop{Tr} [T_R^2 T_j] f_R^2 f_i \nonumber \\
      &   & + \frac{1}{2} \sum_{ij}  \tmop{Tr} [T_R T_i T_j] f_R f_i f_j + \frac{1}{6} \sum_{ijk} \tmop{Tr} [T_i T_j T_k] f_i f_j f_k \nonumber \\
      &   & - \frac{1}{24} p_1 (M) \tmop{Tr} [T_R] f_R - \frac{1}{24} p_1 (M) \sum_i \tmop{Tr} [T_i] f_i
  \label{eq:I6-2}
\end{eqnarray}
where $\tmop{Tr}[T_{A_1} \cdots T_{A_l}] \equiv k_{A_1 \dots A_l}$ are the $l$--degree 't Hooft anomaly coefficients of the 4d theory.
\\

Having compactified the theory on $\Sigma$ it is natural to identify the anomaly polynomial of the corresponding two--dimensional theory with the 
expression obtained by integrating $I_6$ on the Riemann surface.
The result of the integration is 
\begin{equation}
 \label{eq:integratedI6}
  \int_{\Sigma} I_6 = \nu \left[
                                \frac{\tmop{Tr}[T_R^2 T]}{2} f_R^2
                                + \sum_{i} \tmop{Tr}[T_R T_i T] f_R f_i
                                + \sum_{ij} \frac{\tmop{Tr}[T_i T_j T]}{2} f_i f_j
                                - \frac{k}{24} p_1 (\mathbbm{T})
                          \right]
\end{equation}
which can be compared to the general formula for the anomaly polynomial in $2d$
\begin{eqnarray}
  I_4 & = & \tmop{ch}_2 (E) - \frac{1}{24} p_1 (\mathbbm{T}) \tmop{ch}_0 (E) \nonumber\\
      & = & \frac{k_{RR}}{2} f_R^2 + \sum_{i} k_{Ri} f_R f_i + \sum_{ij} \frac{k_{ij}}{2} f_i f_j - \frac{k}{24} p_1 (\mathbbm{T})  
\end{eqnarray}
leading to the following identities
\begin{eqnarray}
 \label{eq:relations}
 k_{RR} & = & \nu \tmop{Tr} [T_R^2 T] \nonumber \\
 k_{Ri} & = & \nu \tmop{Tr} [T_R T_i T] \nonumber \\
 k_{ij} & = & \nu \tmop{Tr} [T_i T_j T] \nonumber \\
 k      & = & \nu \tmop{Tr} [T]
\end{eqnarray}
where $\nu$ is defined in (\ref{eq:nu}). We note that (\ref{eq:relations}) relates 4d 't Hooft anomaly coefficients on the right hand side with 2d anomaly
coefficients, $k_{AB}\equiv\tmop{Tr}[t_A t_B]$, on the left hand side.    

As already mentioned, when we flow to two dimensions the generator $T_R$ corresponding to a trial four dimensional R--symmetry can
mix with the other global $U(1)$'s to give rise to the exact two dimensional R--symmetry.
Therefore, reinterpreting equation (\ref{eq:relations}) in a two--dimensional language, requires substituting
the generator $T_R$ with (\ref{eq:mixR-symmFlav}). Explicitly, we find 
\begin{equation}
  k_{RR}^{trial} = \nu \left[ \xi_i \xi_j \left(\frac{\kappa}{2} k_{ijR} \!+\! b_k k_{ijk} \right)
  \!+\! 2 \xi_i \left( \frac{\kappa}{2} k_{RiR} \!+\! b_j k_{Rij} \right) \!+\! \left( \frac{\kappa}{2} k_{RRR}
  \!+\! b_i k_{RRi} \right) \right]
\end{equation}
\begin{equation}
  k_{Ri}^{trial} = \nu \left[ \left( \frac{\kappa}{2} k_{ijR} + b_k k_{ijk} \right) \xi_j
  + \left( \frac{\kappa}{2} k_{RiR} + b_j k_{Rij} \right) \right]
\end{equation}
\begin{equation}
\label{eq:kij}
  k_{ij} = \nu  \left( \frac{\kappa}{2} k_{ijR} + b_k k_{ijk} \right)
\end{equation}
\begin{equation}
  k = \nu  \left( \frac{\kappa}{2} k_R + b_i k_i \right)
\end{equation}
The mixing parameters $\xi_i$ are now determined by extremizing the trial central charge $c_r^{trial} = - 3 k_{RR}^{trial}$ (a sign appears, due to our choice of 2d chirality matrix $\g_{01}$, see Table \ref{tab:N=1})
\begin{equation}
  \label{eq:max} 0 = \frac{\partial c_r^{trial}}{\partial \xi_i} = -6 k_{Ri}^{trial} 
\end{equation}
which implies
\begin{equation}
 \label{eq:maxeq}
 k_{i \nocomma j} \xi_j + \nu \left(\frac{\kappa}{2}k_{RRi} + b_j k_{Rij}\right) = 0
\end{equation}
Equation (\ref{eq:maxeq}) can be solved by inverting the matrix $k_{ij}$, provided that it has
non-vanishing determinant. The expression for the extremized central charge is finally given by 
\begin{eqnarray}
\label{eq:krrex}
  c_r & = &
  3\nu^2 \left( \frac{\kappa}{2} k_{RRi} + b_k k_{Rki} \right) k^{-1}_{ij}
  \left( \frac{\kappa}{2} k_{RRj} + b_l k_{Rlj} \right) -3 \nu \left( \frac{\kappa}{2} k_{RRR} + b_m k_{RRm} \right)
  \nonumber \\
\end{eqnarray}
in terms of the anomaly coefficients of the original four dimensional SCFT. 

We note that eq. (\ref{eq:maxeq}) determines the coefficients $\xi_i$ 
at the (possible) 2d superconformal fixed point,
giving raise to the exact 2d R-current once plugged in (\ref{eq:mixR-symmFlav}). These coefficients
may differ from the ones appearing in the 4d exact R-current.
There are abelian currents that do not mix in 4d but their mixing in 2d is in general non-vanishing.
This is for example the case of the baryonic symmetries in the $Y^{pq}$ models discussed in \cite{Benini:2015bwz}.
 
\section{Further directions}
\label{sec:conclusions}

We conclude our analysis by discussing some open questions and future lines of research.
A first generalization of the program of constructing 2d SCFTs from four dimensions consists of decorating the Riemann surfaces
discussed here with punctures. 
A possible way to study such a problem consists of exploiting the doubling trick
discussed in \cite{Nagasaki:2014xya,Nagasaki:2015xsa}. In this case one can gain information on the effective number of 2d chiral fermions
by gluing a Riemann surface with a copy of itself (with the opposite orientation), thus
obtaining a closed surface.

One can apply our results to classes of 4d SCFTs with a gravitational dual.
For example one can consider theories associated to D3 branes probing the
tip of three dimensional Calabi--Yau cones.
The analysis of such models was initiated in \cite{Benini:2015bwz}, for the 
infinite $Y^{pq}$ family of \cite{Benvenuti:2004dy}.
Such theories are characterized by the presence of a
$SU(2) \times U(1)$ mesonic flavor symmetry and a $U(1)$ baryonic symmetry.
The baryonic symmetry does not mix with the 4d R--current, but it has been observed that
this mixing is non--trivial once the theory is reduced to 2d.
For more general quivers 
the gauge group is a product of $U(N)_i$ factors.
In the IR the $U(1)_i \subset U(N)_i$ are free and decouple. 
The non anomalous  combinations of these $U(1)$s are the baryonic symmetries.
While in the $Y^{pq}$ case there is just a single baryonic symmetry, in other cases
one can have a richer structure.
The formalism developed in section \ref{sec:central} is necessary for 
extending the analysis to such families.

One can also study the problem from the AdS dual setup along the lines of  \cite{Benini:2015bwz}, 
reconstructing the central charge from the gravitational perspective.
The solution in this case should correspond to D1 branes probing a type IIB
warped AdS$_3 \times_\omega \mathcal{M}_7$ geometry,
where $\mathcal{M}_7$ represents (locally) a $U(1)$ bundle
over a 6d Kahler manifold.
It should be possible to formulate the central charge and its extremization in terms of the
volumes of $\mathcal{M}_7$, in the spirit of \cite{Martelli:2005tp}.

It should be also possible to study models arising from the compactification of 6d theories,
such as class S theories \cite{Gaiotto:2009we} or theories with lower supersymmetry, as the $S_k$ models \cite{Gaiotto:2015usa} or the models of \cite{Bah:2012dg}.
Also the analysis of $\mathcal{N}=3$ theories may be an interesting problem, especially because the 
central charges $a$ and $c$ can be computed along the lines of \cite{Aharony:2016kai}.
The analysis of the gravitational dual mechanism of the topological twist in this case
can be performed by studying the consistent truncation of \cite{Ferrara:1998zt} in gauged supergravity.
It would require to further truncate the $\mathcal{N}=6$ theory to an $\mathcal{N}=2$ subsector once the fluxes are turned on.
In such a case it might be possible to compare the field theory and the supergravity 
 results.

\section*{Acknowledgements}

We thank N.~Bobev,  S.~Schafer-Nameki, J.~van Muiden and A.~Van Proeyen   for useful discussions. 
The work of A.A.  is supported by the Swiss National Science Foundation
(snf) under grant number pp00p2$_{-}$157571/1. This work has been supported in part by Italian Ministero dell'Istruzione, Universit\`a e Ricerca (MIUR), Istituto
Nazionale di Fisica Nucleare (INFN) through the ``Gauge Theories, Strings, Supergravity'' (GSS) research project and MPNS--COST Action MP1210 ``The String Theory Universe''. 
 
\appendix

\section{The anomaly polynomial}
\label{app:A}

In this section we briefly review the general formalism of the anomaly polynomial that has been
used in section \ref{sec:central}. 
We refer the reader to the original paper \cite{AlvarezGaume:1984dr} for further details (see also \cite{Bilal:2008qx} for a review).

We begin by recalling the Atiyah--Singer index theorem for the Dirac operator on a compact manifold.
Let $M$ be a compact closed manifold of even dimension $2l$ and $E$ a smooth complex vector bundle over it, associated to some representation of a Lie group $G$. If $M$ is a spin manifold, we can define fermionic fields as sections of the spinor bundle $\mathcal{S}(M)=\mathcal{S}^+\oplus\mathcal{S}^-$, where $\mathcal{S}^\pm$ are the two chiral irreducible spinor representations of the spin group of $M$. A chiral fermion field charged under $G$ is then described by a section of the bundle $\mathcal{S}^\pm\otimes E$.

The gamma matrices $\g_a$ act on spinors $\mathcal{S}^\pm$ exchanging their chirality, hence the Dirac operator $\slashed{\mathcal{D}}=\g^\mu\mathcal{D}_\mu$ can be represented by the off--diagonal 2--by--2 matrix
\begin{equation}
 \label{eq:diracop}
 \slashed{\mathcal{D}}=\left[\begin{array}{cc}0&\slashed{\mathcal{D}}^-\\\slashed{\mathcal{D}}^+&0\end{array}\right]
\end{equation}
where the operators $\slashed{\mathcal{D}}^\pm:\mathcal{S}^\pm\otimes E\rightarrow\mathcal{S}^\mp\otimes E$ are the adjoints of each others. The Atiyah--Singer index theorem then states that
\begin{equation}
 \label{eq:ASthm}
 \tmop{index} (\slashed{\mathcal{D}}^+) \equiv \dim \ker \slashed{\mathcal{D}}^+ - \dim \ker \slashed{\mathcal{D}}^-
                                        = \int_M \hat{A} (M) \tmop{ch}(E)  \Big|_{2l}
\end{equation}
where $\hat{A} (M)$ is the so called $A$--roof genus of the tangent bundle of $M$
\begin{equation}
\label{roofgenus}
  \hat{A} (M) = 1 - \frac{1}{24} p_1 (M) +
  \frac{1}{5760}  [7 p_1 (M)^2 - 4 p_2 (M)] + \ldots
\end{equation}
expressed in terms of the Pontryagin classes $p_i(M)\in H^{4i}(M,\mathbbm{Z})$, while $\tmop{ch}(E)$
is the Chern character of the bundle $E$. 

In the particular case of $G = U(1)$, all its irreducible representations
are one--dimensional and the bundle $E$ can be decomposed as a Whitney sum of line bundles
$E=\mathcal{L}^{(1)}\oplus\cdots\oplus\mathcal{L}^{(n)}$, one for each representation (particle species).
It follows that each bundle $\mathcal{L}^{(r)}$ is defined, up to isomorphisms, by its first Chern class
\begin{equation}
 \label{eq:chernclass}
 c_1(\mathcal{L}^{(r)})=\frac{1}{2\pi}\left[R(A^{(r)})\right]\in H^2(M,\mathbbm{Z})
\end{equation}
where $\left[R(A^{(r)})\right]$ is the cohomology class of the curvature of the associated abelian connection $A^{(r)}_\mu$.
In this case the Chern character is defined additively as
\begin{equation}
 \label{eq:chernchar}
 \tmop{ch}(E)=\sum_{k=0}^\infty\tmop{ch}_k(E)=\sum_{k=0}^\infty \frac{1}{k!}\left[c_1(\mathcal{L}^{(1)})^k+\cdots+c_1(\mathcal{L}^{(n)})^k\right]
\end{equation}
Since each line bundle $\mathcal{L}^{(r)}$ is associated to a unitary one--dimensional representation of
integer charge $q^{(r)}$, we can equivalently describe the bundle $E$ as follows. If we define $f_G$ to
be the first Chern class of the line bundle of unit charge\footnote{Note that a principal $U(1)$
bundle and the associated line bundle of charge 1 have the same first Chern class.}, then for each
$\mathcal{L}^{(r)}$ we can write
\begin{equation}
 \label{eq:chernclass2}
 c_1(\mathcal{L}^{(r)})=q^{(r)} f_G
\end{equation}
from which we obtain
\begin{equation}
 \label{eq:chernclass2}
 c_1(E)=\tmop{Tr}[T_G] f_G
\end{equation}
where we assembled all the charges $q^{(r)}$ into the diagonal matrix $T_G$, which now represents the Lie
algebra part of the connection on $E$. Using this redefinitions, the Chern character (\ref{eq:chernchar}) is
\begin{equation}
 \label{eq:chernchar2}
 \tmop{ch}(E)=\sum_{k=0}^\infty \frac{\tmop{Tr}[T_G^k]}{k!}f_G^k
\end{equation}
More generally, for a family of $n$ fermions charged under $m$ abelian symmetries $G=\prod_{i=1}^m U(1)_i$, we have to consider the bundle
\begin{equation}
 \label{eq:productbundle}
 E = \bigoplus_{r=1}^n \mathcal{L}^{(r)} \qquad \text{with} \qquad \mathcal{L}^{(r)}=\mathcal{L}^{(r)}_1 \otimes\cdots\otimes \mathcal{L}^{(r)}_m
\end{equation}
where each $\mathcal{L}^{(r)}$ is a tensor product representation for the group $G$, labelled by the
set of charges $(q^{(r)}_1,\dots,q^{(r)}_m)$. If, as before, we define $f_i$ to be the first Chern class of the
line bundle of unit charge for the $U(1)_i$ symmetry, we can write
\begin{eqnarray}
 \label{eq:chernchar3}
 \tmop{ch}_k(E) & = & \frac{1}{k!} \left[c_1(\mathcal{L}^{(1)})^k+\cdots+c_1(\mathcal{L}^{(n)})^k\right]\nonumber\\
                & = & \frac{1}{k!} \sum_{r=1}^n \left(\sum_{i=1}^m q^{(r)}_i f_i\right)^k\nonumber\\
                & = & \frac{1}{k!} \sum_{i_1\cdots i_k}^m \left(\sum_{r=1}^n q^{(r)}_{i_1} \cdots q^{(r)}_{i_k} \right) f_{i_1}\cdots f_{i_k} \nonumber\\
                & = & \frac{1}{k!} \sum_{i_1\cdots i_k}^m \tmop{Tr}\left[T_{i_1} \cdots T_{i_k} \right] f_{i_1}\cdots f_{i_k}
\end{eqnarray}
where we used the property of the first Chern class, $c_1(\mathcal{L}^{(r)}_i\otimes\mathcal{L}^{(r)}_j)
=c_1(\mathcal{L}^{(r)}_i)+c_1(\mathcal{L}^{(r)}_j)$. The diagonal matrices $T_i$ can be taken to be
the hermitian generators of the $U(1)_i$ symmetries, written in the representation associated to $E$.
\\

The physical interpretation of the index of $\slashed{\mathcal{D}}^+$ is that of a chiral anomaly for
the effective action of a massless chiral fermion $\psi\in\g(\mathcal{S}^+\otimes E)$:
\begin{equation}
 \mathe^{\mathi W }=   \int d \psi d
  \bar{\psi} \, \mathe^{ \int \bar{\psi}\, \mathi \slashed{\mathcal{D}}^+ \psi}
\end{equation}
In fact, under a chiral rotation the fermionic path integral picks up a non--zero phase 
\begin{equation}
  \delta W =   -2 \, \tmop{index} (\slashed{\mathcal{D}}^+)
\end{equation}
proportional to the index of the Dirac operator. 

Since the anomaly for a negative chirality
fermion is minus that of a positive chirality fermion
\begin{equation}
 \label{eq:indexofDminus}
 \tmop{index}(\slashed{\mathcal{D}}^-) \equiv \dim \ker \slashed{\mathcal{D}}^- - \dim \ker \slashed{\mathcal{D}}^+=-\tmop{index} (\slashed{\mathcal{D}}^+)
\end{equation}
in a theory with many fermions of both chiralities, the total chiral anomaly is given
by the sum of the anomalies of the positive--chirality fermions minus the sum of the anomalies of the
negative--chirality ones.

Finally, in \cite{AlvarezGaume:1984dr} it was shown that one can construct a Dirac operator in $2l+2$
dimensions in such a way that its index reproduces the gauge anomaly for a charged chiral fermion in
$2l$ dimensions. The corresponding index density is a $(2l+2)$--form
\begin{equation}
\label{anomalypol}
I_{2l+2} \equiv \hat{A} (M) \tmop{ch} (E) \Big|_{2l+2}
\end{equation}
which is called the anomaly polynomial.

\section{Supersymmetry variations of the auxiliary fields in $\mathcal{N}=3, 4$ SCFTs}
\label{app:B}

In this section we show that it is always possible to satisfy the conditions $\delta \chi_{IJ}=0$, $\delta \zeta^I =0$ in (\ref{eq:N=3varchi}, \ref{eq:N=3varzeta}) and $\delta \chi^{IJ}_K=0$ in (\ref{eq:N=4varchi}) by assigning a non--vanishing value to the background auxiliary field $D^I_J$ and $D^{IJ}_{KL}$, respectively. 

Note to the reader: in this section we do not assume Einstein summation notation for repeated R--symmetry indices.  

We begin by considering the ${\mathcal N}=4$ case. Since we gauge the background R--symmetry along a subgroup
of the Cartan of $SU(4)$, the curvature $R(V)^I_J \equiv R(V^A) (\mathi \lambda_A)^I_J$ is diagonal in the adjoint $(I,J)$ indices. As a consequence, the Killing spinor equations (\ref{eq:killingN=4})
split into a set of four decoupled equations for $\varepsilon^I$, $I=1, \dots ,4$. Non--trivial $\varepsilon^I$ solutions correspond to the preserved supersymmetries, whereas the rest of the components are set to zero.  

Having this in mind, we now discuss the condition $\delta \chi^{IJ}_K=0$, where the variation is generated by the preserved supercharges. Three possible cases can arise. 

If $K\neq I,J$ from (\ref{eq:N=4varchi}) we immediately find  
\begin{equation}
 \label{eq:B1}
 \delta \chi^{IJ}_K = \frac12 \sum_L D^{IJ}_{KL}\varepsilon^L=0
\end{equation}
that can be trivially solved by setting the corresponding $D^{IJ}_{KL}$ components to zero.

The second case corresponds to $K=I\neq J$ with non--vanishing $\varepsilon^I$ and $\varepsilon^J$. Restricting as usual to the positive chirality transformation, the $\chi^{I  J}_I $ variation reads
\begin{equation}
 \label{eq:B2}
 \delta \chi^{I  J}_I =   \frac12 D^{IJ}_{IJ} \epsilon^J
                        - \frac14\g^{ab}R_{ab}(V)^I_I \epsilon^J -\frac{1}{12} \g^{ab}R_{ab}(V)^J_J \epsilon^J =0
\end{equation}
where we chose $D^{IJ}_{IL}$ to be diagonal in the $J,L$ indices. 

After twisted compactification the $\epsilon^J$ spinors decompose as $\mathi \g_{23}$ eigenvectors
and we write $\mathi \g_{23} \epsilon^J = s_J \epsilon^J$ with eigenvalue $s_J = \pm 1$ according to the 2d chirality of the spinor. 
Using equation (\ref{eq:killingN=4})
\begin{equation}
 \label{eq:B3}
 \frac12 R_{\mu\nu}(\omega^{23})\g_{23}\epsilon^J = R_{\mu\nu}(V)^J_J \epsilon^J \qquad {\rm with} \qquad R_{23}(\omega^{23}) = \kappa \Omega_{23}
\end{equation}
we eventually find 
\begin{equation}
 \label{eq:B4}
 \delta\chi^{IJ}_I = \left[ \frac12 D^{IJ}_{IJ} + \frac{\kappa}{ 4}|e|\Omega_{23}   \left(s_I s_J + \frac13 \right) \right]\varepsilon^J=0
 \end{equation}
If $\kappa = 0$ this equation is easily satisfied by $D^{IJ}_{IJ} =0$. If $\kappa \neq 0$, from Table \ref{TableN4} it turns out that for each fixed $\epsilon^J$ solution only one chirality is present and (\ref{eq:B4}) can be always satisfied by an appropriate choice of the $D^{IJ}_{IJ}$ components.

Finally, if $\varepsilon^J$ is a Killing spinor but $\varepsilon^I$ is not, the $\chi^{IJ}_I$ variations do not vanish in general.
However it is possible to show with a case by case analysis that these components always decouple from the representation of the supersymmetry algebra and they are not relevant for the counting of the supersymmetries.
\vskip 5pt
For the $\mathcal{N}=3$ case, solutions to (\ref{eq:N=3varchi}, \ref{eq:N=3varzeta}) can be derived from the general $\mathcal{N}=4$ solution 
by recalling that the fermionic auxiliary components of the $\mathcal{N}=3$ Weyl multiplet can be obtained from the $\mathcal{N}=4$ ones according to the following decomposition \cite{vanMuiden:2017qsh}
\begin{equation}
 \chi_{(KL)} + \sum_M\varepsilon_{KLM} \zeta^M \equiv \sum_{IJ}\frac{1}{2} \varepsilon_{LIJ4} \,
 \chi^{IJ}_K\, , \quad \quad D^M_N \equiv \sum_{IJKL}\frac{1}{4} \varepsilon^{MKL4}
  \varepsilon_{NIJ4} D^{IJ}_{KL}
\end{equation}

Therefore, exploiting the previous results, we conclude that also in the $\mathcal{N}=3$ case it is always possible to choose a non--vanishing $D^M_N$ background that makes the supersymmetry variations 
$\delta\chi_{IJ}$, $\delta \zeta^I$ vanishing.

\bibliographystyle{JHEP}
\bibliography{References}
\end{document}